\documentclass[superscriptaddress,twocolumn,prb]{revtex4-2}

\usepackage{amsmath,amssymb,amsfonts}
\usepackage[version=4]{mhchem}
\usepackage{bm}
\usepackage{booktabs}
\usepackage{graphicx}
\usepackage{bm}
\usepackage{xcolor}
\usepackage{siunitx}
\usepackage{lipsum}
\usepackage{changes}
\usepackage[colorlinks=True,linkcolor=blue,anchorcolor=red,citecolor=blue,CJKbookmarks=true,urlcolor=blue]{hyperref}

\begin{document}


\title{Interlayer excitons in double-layer transition metal dichalcogenides quantum dots}
\author{Xiang Liu} 
\author{Zheng Tao} 
\author{Wenchen Luo} \email{luo.wenchen@csu.edu.cn}
\affiliation{School of Physics, Central South University, Changsha, China 410083}
\author{Tapash Chakraborty} 
\affiliation{Department of Physics and Astronomy, University of Manitoba, Winnipeg,
Canada R3T 2N2}
\date{\today}

\begin{abstract}
Various properties of interlayer excitons in double-layer transition metal dichalcogenides quantum dots are  {analyzed} using a low-energy effective Hamiltonian with Coulomb interaction.
We solve the single-particle Hamiltonian with and without a magnetic field  analytically, then present the electron-hole pairing features of interlayer exciton by employing  {the exact diagonalization technique}, where the electron and hole  {are located} in two layers respectively. 
In a magnetic field, the Landau level gap, as well as the electron-hole separation of an exciton varies non-monotonously as the interlayer distance increases, attributed to the pseudospin-orbit coupling which also leads to the emergence of topological non-trivial pseudospin textures in the exciton states. 
We examine the influence of different materials in quantum dots stacking on the exciton states, comparing their impact to variations in 
layer distances and quantum dot sizes. We further explore two interacting interlayer excitons numerically. The binding energy is significantly enhanced by the exchange interaction when the two electrons have different spins.
The optical absorption spectra from the ground state to low-lying excited states reveal distinct behaviors for different interlayer excitons, which can be utilized to distinguish the spin of electrons in excitons. 
Our results highlight the potential for controlling interlayer excitons and applications of optical devices in a magnetic field and tunable layer distance.     
\end{abstract}

\maketitle
\section{INTRODUCTION}
Interlayer exciton form when electron and hole with strong  {mutual} Coulomb interaction  are localized in different monolayers. Recently, this emerging phenomenon  {has received increasing attention}, prompting extensive theoretical and experimental studies, particularly within transition metal dichalcogenides (TMDs)\cite{palinterlayerex,latini2017interlayer}. TMDs are a novel family of two dimensional (2D) materials with unique optoelectronic properties, and are highly useful in various fields, including optoelectronic device, energy and  {medical} applications\cite{2017app,2018app}. 
These materials, described by the formula \ce{MX2} (M = Mo, W and X = S, Se), are distinguished from graphene by their strong spin-orbit couplings (SOCs) originating from heavy transition metal atoms\cite{yao2012tmd}. In contrast to their bulk counterparts, monolayer TMDs have a direct band gap of about $2$ eV, making them attractive for optical studies and applications. 
Most heterostructures exhibit a type-II band alignment under specific conditions, where the conduction band minimum (CBM) and valence band maximum (VBM) lie in different layers\cite{geim2013van,bandtype,kang2013band}. Alternatively, voltage modulation can also be used to form the type-II band alignment \cite{unuchek2019valley,hajati2023tuning}. These heterostructures facilitate the formation of interlayer excitons and provide a versatile platform for exploring correlated electronic states, paving the way for novel optical and electronic phenomena\cite{zhao2022interlayer}. Recent works on interlayer excitons in TMDs have revealed different behavior from intralayer excitons, particularly the lifetime of interlayer exciton can be up to two orders of magnitude longer than that of intralayer exciton \cite{rodin2013excitonic,gartstein2015exciton,van2018interlayer,wu2019exciton,tuninterlayer2023}. 

Excitons in bilayer two-dimensional electron systems have,  {of late,} garnered significant interest, which supports supercurrent, excitonic superfluidity, novel crystal phases, etc. \cite{su2008make, fu2024bilayer, 2017superfluidity}. 
The heterostructures, including both homobilayers (made from the same TMDs) and heterobilayers (made from different TMDs) \cite{2024multiferroicityTMDS}, additionally hybrid heterostructures combining perovskite and TMDs have recently been fabricated\cite{yuan2023controlled,yuan2024vdw,lu2024van}. These stacking structures have been extensively researched for their optoelectronic properties and applications, which underscores the primary focus of our study.

The confined nanostructures such as quantum dots (QDs) and quantum rings based on semiconductor heterostructures were investigated extensively and were also utilized in optical devices \cite{tapsh1999129,chakraborty2003nanoscopic,tapshringbook,chen2007fock}. 
The optical absorption spectra in a parabolic QD, first studied theoretically in Ref.\cite{tapshexciton} were in excellent agreement with the observed low-energy emission lines in a subsequent experiment \cite{zrenner1994quantum}.
Theories and experiments on electrons in elliptical and circular QDs  have displayed fascinating optical and transport properties influenced by confinement effects, focusing on both single-particle and few-body systems.
Numerous studies on graphene and TMDs demonstrate various techniques for implementing QDs in 2D materials, which paves the way for exploring and applying these artificial atoms with novel electronic and optical properties \cite{mac2022strong,chen2007fock,tapsh2013electron,2022gateQD,js1,js2,shu,luooe} comparing with the QDs in conventional semiconductors.
Electrons trapped in bilayer and trilayer graphene QDs have been studied numerically by using finite element method, where interlayer coupling is considered \cite{tapsh2013electron,peetertrilayercomsol}. The QD located in a monolayer TMDs nanoflake
can also be experimentally achieved and are studied theoretically \cite{kormanyos2014spin,liu2014intervalley,qu2017tunable,qu2016robust,miravet2023interacting,vadymqd,vadym2024ultrafast}.

Motivated by the vast application potential of TMDs, in the present work, we explore interlayer excitons by constructing a double QDs structure based on TMDs, where electrons and holes are located in different layers. Between the two layers, aside from the Coulomb interactions, the particle hopping across different layers is disregarded. The distance of the two layers and an external magnetic field are supposed to be adjustable. We investigate how the layer distance and magnetic field influence the properties of interlayer excitons within the double-layer QDs. Our device based on TMDs would be ideal for studying its optical and transport properties. This system with strong Coulomb interactions are also promising for studying electron-hole pairing and enhancing understanding of correlated electronic states \cite{2023quantWigner, topsuper2024}.

The manuscript is organized as follows. In Sec.~\ref{sec:model}, we present our model for electron-hole pairs in double TMDs QDs, including the Hamiltonian formalism and a review of single-particle solutions within the TMDs QDs. We then introduce the interlayer Coulomb interaction matrix elements used for the numerical analyses. To streamline the discussion, we address the Coulomb interaction matrix elements separately in the absence and presence of an external magnetic field. In Secs.~\ref{sec:exciton}, we employ the exact diagonalization (ED) techniques to calculate the ground-state and low-lying excited energies, electron-hole separation, and optical absorption strengths of interlayer excitons for double TMDs QDs. Additionally, the discussion encompasses the topological properties of interlayer exciton.  In Secs.~\ref{sec:biexciton}, the results of biexciton are discussed, where the exciton-exciton interaction plays an important role. In Sec.~\ref{sec:summary} we make a summary and outlook on the exciton features. 
\section{MODEL HAMILTONIAN} \label{sec:model}

In our model, two monolayer TMDs are positioned in the upper and lower planes, respectively, and are separated by a hexagonal boron nitride (hBN) substrate \cite{hbn71,van2018interlayer}. The two TMDs layers could form either homobilayers or heterobilayers.
We consider the widely adopted low-energy effective model to describe the behavior of electrons near the Fermi surface\cite{yao2012tmd}. Recent studies employ the periodic moir\'e potential in moir\'e superlattices to form quantum dot arrays\cite{2023quantWigner,mac2022strong,luo2023artificial}. However, we construct a double QDs structure confining electrons and holes in different monolayers respectively by using an infinite potential. The advantage of our approach is that it allows for convenient transport measurements and the study of optical properties 
with minimal interference. In this work, we focus on interlayer excitons with long-range Coulomb interaction and ignore the  {hopping} between the layers given that the distance of two layers are relatively large.
The system is described by the Hamiltonian
\begin{equation} \label{eq:Hamiltonian}
H = H_{e} + H_{h}  + H_{eh} + H_{ee} +H_{hh} ,	
\end{equation}
where $ H_{e} $ and $ H_{h} $ respectively correspond to the single-particle Hamiltonian in the upper and lower layer QDs, $ H_{eh} $ is the Coulomb interaction between the electron and the hole located in different layers, $ H_{ee} $ describe the intra-layer electron-electron interaction and $ H_{hh} $ is the intra-layer hole-hole interaction. Without loss of generality, we suppose that electrons are in the upper QD and holes locate in the lower QD.

The Hamiltonian of the single particle in TMDs QD is 
\begin{equation} \label{eq:ham}
H_{\zeta}=\frac{a_{\zeta}t_{\zeta}}{\hbar }H_{\tau} +%
\frac{\Delta_{\zeta}}{2}\sigma _{z}+\frac{\lambda_{\zeta}\tau s}{2}\left( 1-\sigma
_{z}\right) + V(\textbf{r}_{\zeta})\sigma_{z},
\end{equation}
where  {$\zeta=e,h$} denotes the index of electron or hole layer, $ a_\zeta $ is the lattice constant of the TMDs of layer, $ t_\zeta $ is the hopping parameter, $ \tau $ is the valley index that $ \pm 1 $  for $  K  $ and $ K^{\prime} $ valley respectively, $\sigma_z$ is a Pauli matrix representing the sublattice pseudospin, $H_{\tau}=\tau\sigma_{x}\left(p_{x}+eA_{x}\right) + \sigma_{y}\left( p_{y}+eA_{y}\right)$ with vector potential $\mathbf{A}=(A_x,A_y)$ in a perpendicular magnetic field represents the sublattice pseudo SOC, $\Delta $ is the energy gap between valence and conductance bands, $ \lambda $ is a constant coupling spin and valley, $ s = \pm 1 $ for spin up and down acting as a good quantum number respectively, and $\bm r_{12}$ are the position vectors in different layers. We neglect the small Zeeman term in Eq.~\eqref{eq:ham} in our model.

The QD can be produced by cutting a monolayer TMDs nanoflake of radius $R$ or lateral confinement potential on an extended monolayer\cite{kormanyos2014spin,liu2014intervalley,qu2017tunable,
qu2016robust,miravet2023interacting}.
Then we adopt the the infinite mass boundary\cite{berry1987neutrino}
for the QD, which allows us to obtain exact solutions of single-particle wave functions,
\begin{align}
V\left(\textbf{r}\right) =\left\{ 
\begin{array}{c}
0 \\ 
\infty 
\end{array}%
\right. 
\begin{array}{c}
r<R \\ 
r>R%
\end{array} , \label{eq:V_conf}
\end{align} 
To solve the many-particle system, we should first establish the single-particle eigenstates of the non-interacting Hamiltonian as the basis. This Hamiltonian, denoted as $ H_{0} = H_{e} + H_{h}+V$. The solutions of the single-particle  {Hamiltonian} are given in the following.


\subsection{Wave functions without a magnetic field}\label{subsec:nomag}

In the absence of magnetic field, the wave function is an eigenstate of the total angular momentum operator $ J_{tol} = L_{z} + \tau\frac{\hbar\sigma_{z}}{2}$, since it commutes with the Hamiltonian, $\left[J_{tol},H_{\zeta}\right] =0$, where $L_z=xp_y - yp_x$ is the $z$ component of the angular momentum. Given that the spin is a good quantum number, we start form the eigenvalue equation $ H_\zeta \Psi_\zeta = E_\zeta\Psi_\zeta$ for spin $s$ in polar coordinate $(r,\theta)$ with the wave function being a two-component spinor. Considering that in either upper or lower layer, the eigenvalues equations are the same in form. We drop the layer index in this section, unless otherwise specified. The wave function should be labeled by four indices, principal quantum number $n$, angular momentum $j$, valley $\tau$, and spin $s$. For simplicity, the wave function spinor, without the four indices, can be written as 
\begin{equation}
\psi (r,\theta)=\left( 
\begin{array}{c}
e^{i\left( j-\frac{\tau }{2}\right) \theta }\chi_{1}\left( r\right)  \\ 
ie^{i\left( j+\frac{\tau }{2}\right) \theta }\chi_{2}\left( r\right) 
\end{array}%
\right), \label{eq:waf}
\end{equation}
where the total angular momentum quantum number $j=m +\frac{\tau}{2}$ with orbital angular momentum  quantum number $m=0,\pm1,\pm2... $, while  $\chi_{ 1}(r)$ and $\chi_{ 2}(r)$ are the radial wave functions that are the envelope functions containing the principal quantum number $n$. The eigenstate problem can be written as two coupled differential equations for the two spinor components,
\begin{align}
a t \left(\tau \nabla_{r}+\frac{j+\frac{\tau}{2}}{r}\right)\chi_{ 2}(r)&=\left( -\frac{\Delta }{2}+E_j \right)\chi_{1}(r),  \label{eq:sub1}\\
a t \left(\tau \nabla_{r} -\frac{j-\frac{\tau}{2}}{r}\right)\chi_{ 1}(r)
&=\left( \lambda \tau s-\frac{\Delta }{2}-E_j \right)\chi_{ 2}(r),  \label{eq:sub2}
\end{align}
where $ E_{j} $ is the eigenenergy with total angular momentum quantum number $j$. 
Following a series of mathematical steps (See Appendix.~\ref{app:basis} for detailed derivations), we obtain the single-particle wave function,
\begin{equation}
\psi=N\left( 
\begin{array}{c}
e^{i\left(j-\frac{\tau }{2}\right) \theta}\frac{2a t}{2E_j -\Delta}\sqrt{\kappa_j} \varrho J_{(j-\frac{\tau }{2})}\left( 
\sqrt{\kappa_j}r\right)  \\ 
ie^{i\left( j+\frac{\tau }{2}\right) \theta } \varrho J_{(j+\frac{\tau}{2})%
}\left(\sqrt{\kappa_j}r\right) 
\end{array}%
\right),
\end{equation}
where $N$ is normalization coefficient, $J$ is a Bessel function of the first kind and we define $\kappa_j =\left( 2E_j -\Delta \right) \left( 2E_j+\Delta -2\tau s\lambda\right) /4a^{2}t^{2}$. Here, 
$ \varrho $ assumes distinct values corresponding to the various angular momentum quantum states. The infinite mass boundary condition implies that the eigenvalues equations satisfy 
\begin{equation} 
\chi_{2}\left(R\right) =i\tau \chi_{1}\left( R\right)   \label{infinitemass}
\end{equation} 
at the border of the QD \cite{berry1987neutrino,pe2011electronic,egg2011finite,egger2011signatures,
dirac2017excitonic}. Through numerical solutions of the boundary condition in Eq.~\eqref{infinitemass}, one can determine a series of energies at the angular momentum $j$ marked as $E_{n,j}$ labeled by quantum number $n$. Consequently, the energy spectrum of the QD is obtained at zero magnetic field.

\subsection{Wave functions in the presence of a magnetic field}\label{sec:mag}

We continue to solve the wave functions in a homogeneous perpendicular magnetic field $B$. In this case, the Hamiltonian also takes the form of Eq.~\eqref{eq:ham}. We derive a pair of coupled differential equations for a two component spinors in the presence of a magnetic field,
\begin{subequations}
\begin{align}
\left( \tau \nabla _{\rho }+\frac{j+\frac{\tau }{2}}{\rho }+\beta \rho
\right) \chi _{2}\left( \rho \right)&=\left( \varepsilon -\frac{1}{2}%
\delta \right) \chi _{1}\left( \rho \right) \label{eq:submag1}  \\
\left( \tau \nabla _{\rho }-\frac{j-\frac{\tau }{2}}{\rho }-\beta \rho
\right) \chi _{1}\left( \rho \right) &=\left( \Lambda \tau s-\frac{1}{2}\delta -\varepsilon \right) \chi _{2}\left( \rho \right), \label{eq:submag2}
\end{align}
\label{eq:diff_mag}
\end{subequations}
where we define dimensionless parameter $\rho=\frac{r}{a}$, $\beta=\frac{eB}{2\hbar}a^{2}$, $\varepsilon=\frac{E}{t}$, $\delta=\frac{\Delta}{t}$ and $\Lambda=\frac{\lambda }{t}$.  Using the same steps as mentioned in last subsection, we substitute Eq.~\eqref{eq:submag2} into Eq.~\eqref{eq:submag1}, 
and make an ansatz $ \chi_{1}\left(\rho\right) =\rho ^{\left\vert j-\frac{\tau }{2}\right\vert }\exp \left( -\frac{\beta \rho ^{2}}{2}\right) \chi_{0}\left( \rho \right)$, then a confluent hypergeometric equation is obtained
\begin{equation}
x\nabla _{x}^{2}\chi _{0}\left( x\right) +\left( b-x\right) \nabla _{x}\chi
_{0}\left( x\right) -\alpha \chi _{0}\left( x\right)=0,
\end{equation}
where we define $ x=\beta\rho^{2} $, $ b=\left\vert j-\frac{\tau }{2}\right\vert +1$ and 
\begin{equation}
\alpha =\frac{1}{2}\left( j+\frac{\tau }{2}+b\right) -\frac{\left(
\varepsilon +\frac{1}{2}\delta -\Lambda \tau s\right) \left( \varepsilon -%
\frac{1}{2}\delta \right) }{4\beta}. 
\end{equation}
After some algebraic operations, we obtain the single-particle eigenstate in the presence of an external magnetic field
\begin{equation}
\psi =N\left( 
\begin{array}{c}
e^{i\left( j-\frac{\tau }{2}\right) \theta }\rho ^{\left\vert j-\frac{\tau }{%
2}\right\vert }\exp \left( -\frac{\beta \rho ^{2}}{2}\right) \left.
_{1}F_{1}\right. \left(\alpha,b,\beta \rho ^{2}\right)  \\ 
ie^{i\left( j+\frac{\tau }{2}\right) \theta }\rho ^{\left\vert j-\frac{\tau 
}{2}\right\vert -1}\exp \left( -\frac{\beta \rho ^{2}}{2}\right) \gamma \Gamma
\left( \rho \right) 
\end{array}%
\right),
\end{equation}
where $N$ is the normalization constant and $ \left. _{1}F_{1}\right.(\alpha,b,r) $  is the confluent hypergeometric function. Here, we define $ \Gamma \left( \rho \right) = \xi \left. _{1}F_{1}\right. \left(\alpha,b,\beta \rho^{2}\right) +\tau \frac{2\alpha}{b}\beta \rho ^{2}\left. _{1}F_{1}\right. \left(\alpha+1,b+1,\beta \rho ^{2}\right)$ with $ \xi =\tau \left\vert j-\frac{\tau }{2}\right\vert -j+\frac{\tau }{2}-\left(\tau +1\right) \beta \rho ^{2} $ and $ \gamma=-(\varepsilon +\frac{1}{2}\delta -\Lambda \tau s)^{-1} $. Again, using the infinite mass boundary condition in Eq. (\ref{infinitemass}), we can numerically determine the energy spectrum \cite{berry1987neutrino,chen2007fock,pe2011electronic,egg2011finite,egger2011signatures,dirac2017excitonic}. 

\subsection{Many-body Hamiltonian}
Once the single-particle wave functions are obtained, we can construct the generic many-body basis as $|i\rangle \equiv |i_{e_1},\ldots i_{e_{N_e}}; i_{h_1},\ldots, i_{h_{N_h}} \rangle$ which contains $N_e$ electrons and $N_h$ holes. Each $i_{e_k}$ and $i_{h_k}$ is a collective index containing the indices $n_{e_k},j_{e_k},s_{e_k}$ for electron and indices $n_{h_k},j_{h_k}$ for hole, respectively. For simplicity, in this work, we constrain ourselves to the case of one electron and one hole (one exciton), and the case of two electrons and two holes (two excitons). Without loss of generality, the electrons and holes locate in the upper and lower layers, respectively. Considering that the two conductance bands with different spins are close,  here the electron state is associated with the spin index. In the second quantization, the many-body Hamiltonian is given by

\begin{eqnarray}
H&=&\sum_{s,n,j}  E^e_{j,n,s} c^\dag_{j,n,s} c_{j,n,s}+\sum_{n,j} E^h_{j,n} d^\dag_{j,n} d_{j,n}  \notag \\
&+&H_{ee}+H_{hh}+H_{eh},   \label{many-bodyh}\\
H_{ee}&=& \frac{1}{2} \sum_{s,s'}\sum_{n_{1},\ldots n_{4}} \sum_{j_{1},\ldots j_{4}} V^{(ee) n_1,n_2,n_3,n_4}_{j_{1},j_{2},j_{3},j_{4}}  \notag \\
&\times& c^\dag_{j_{1},n_1,s}c^\dag_{j_{2},n_2,s'}
c_{j_{3},n_3,s'} c_{j_{4},n_4,s},   \\
H_{hh}&=& \frac{1}{2} \sum_{n_{1},\ldots n_{4}} \sum_{j_{1},\ldots j_{4}} V^{(hh) n_1,n_2,n_3,n_4}_{j_{1},j_{2},j_{3},j_{4}} \notag \\
&\times& d^\dag_{j_{1},n_1}d^\dag_{j_{2},n_2} d_{j_{3},n_3} d_{j_{4},n_4},  \\
H_{eh} &=& - \sum_{s}\sum_{n_{1},\ldots n_{4}} \sum_{j_{1},\ldots j_{4}}  V^{(eh) n_1,n_2,n_3,n_4}_{j_{1},j_{2},j_{3},j_{4}}  \notag \\
&\times& c^\dag_{j_{1},n_1,s} d^\dag_{j_{2},n_2} d_{j_{3},n_3} c_{j_{4},n_4,s},  \label{H_hh}
\end{eqnarray}
where $c_{j,n,s}$ and $c^\dag_{j,n,s}$ are the annihilate and creation operators for electron with spin $s$, principal quantum number $n$ and angular momentum $j$, $E^e_{j,n,s}$ is the non-interacting energy of the electron state, $d_{j,n}$ and $d^\dag_{j,n}$ are the operators for hole, and $E^h_{j,n}$ is the energy of a hole state. $H_{ee},H_{hh}$ and $H_{eh}$ are the electron-electron, hole-hole and electron-hole Coulomb interaction, respectively, where the Coulomb interaction matrix elements $V^{(ee) n_1,n_2,n_3,n_4}_{j_{1},j_{2},j_{3},j_{4}}, V^{(hh) n_1,n_2,n_3,n_4}_{j_{1},j_{2},j_{3},j_{4}}$ and $V^{(eh) n_1,n_2,n_3,n_4}_{j_{1},j_{2},j_{3},j_{4}} $ are given in Appendix \ref{app:cou_mat} in detail. 

If there are only one electron and one hole in the system, then the Hamiltonian in Eq.~\eqref{many-bodyh} should exclude $H_{ee}$ and $H_{hh}$.
In our numerical calculation, we adopt  {the ED scheme mentioned above} to solve the many-body Hamiltonian in Eq.~\eqref{many-bodyh}. Then the energy spectra, electron-hole separation, pseudospin textures and the light absorption of the excitons can be studied. Considering the numerical complexity, we set an energy truncation of $50$ meV for electron or hole states in QDs with radii $20$ and $30$ nm, when constructing the Hilbert space of the many-body states. For the radius of the QD being $10$ nm (in the two-exciton case we consider smaller QDs for saving computations), the truncation is increased to $100$ meV. 

\begin{table} 
\caption{The parameters of two materials \ce{MoS2} and WS$_2$, including lattice constants $a$, effective hopping integrals $t$, band gaps  $\Delta $, and SOC constant $ \lambda $, are typically obtained through fitting first-principles band structure calculations. The unit is nm for length and meV for energy. }\label{tab:tabpara}
\begin{ruledtabular}
\begin{tabular}{ccccccccc}
Material	& $a$  & $\Delta$ & $t$ & $\lambda$  \\ \hline
MoS$_2$  & 0.3193 & 1660 & 1100 & 75  \\
WS$_2$   & 0.3197 & 1790 & 1370 & 230 \\
\end{tabular}
\end{ruledtabular}
\end{table}

\section{RESULTS ON a single exciton} \label{sec:exciton}

We conducted numerical studies on TMD vdW heterostructures, 
and heterobilayers.

Although there are numerous interesting 2D TMDs materials\cite{zhou2023controllable,zhou2024stable}, such as \ce{MoTe2} with electronic and structural phase transitions and bulk \ce{ReS2} behaving as decoupled monolayers\cite{MoTe2phase,bulkReS2}, 
we here specifically examine  {the} exemplary \ce{MoS2}/\ce{MoS2} homobilayer QDs and \ce{MoS2}/\ce{WS2} heterobilayer QDs without loss of generality. The TMDs family is vast, yet many members can be described using similar Hamiltonians. 

Other systems could be analyzed in the same method, potentially yielding similar results. The parameters for \ce{MoS2} and \ce{WS2} used in numerical calculations are listed in Table~\ref{tab:tabpara}\cite{yao2012tmd}. For simplicity, we consistently consider the electron and hole within a single valley throughout our analysis, which is deemed justifiable as, the inter-valley interaction is negligible due to the large kinetic difference between two valleys.

 {Additionally, }
achieving valley polarization through optical techniques is feasible in practice \cite{vadymqd,vadym2024ultrafast}. 
We further restrict the hole states with spin-up and consider the electron states of spin-up and spin-down separately in the conduction band. It is motivated by the observation of a large splitting at the valence band edge due to sublattice pseudo SOC.

\subsection{Single-particle energy spectra in TMDs QDs}

\begin{figure}[htp]
\centering
\includegraphics[width=8.4cm]{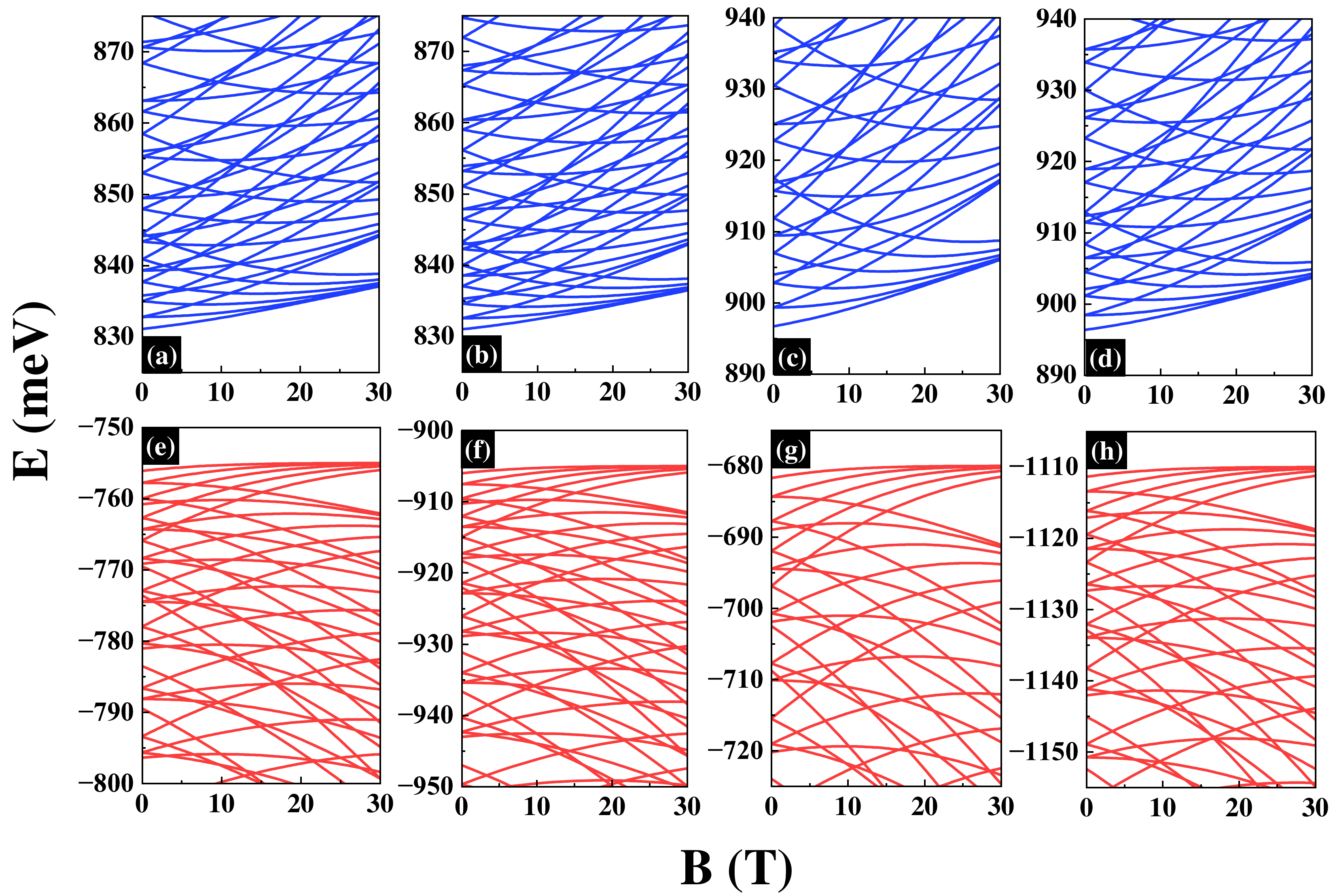}
\caption{The low-lying energy spectra of a single electron and a single hole in TMDs QDs with radius $R = 20$ nm vary with perpendicular magnetic fields. For simplicity, only the first five levels of a principal quantum number are displayed. 
The energy spectra are provided for different TMDs with different spins: (a) \ce{MoS2}, the conductance band with spin $ s = 1 $; (b) \ce{MoS2}, conductance band with $ s = -1 $; (c) \ce{WS2} conductance band with $ s = 1 $; (d) \ce{WS2} conductance band with $ s = -1 $; (e) \ce{MoS2}, valance band with $ s = 1$ (f) \ce{MoS2}, valance band with $ s = -1 $ which is much  {farther} away from the Fermi surface; (g) \ce{WS2}, valance band with $ s = 1 $; (h) \ce{WS2}, valance band with $ s = -1 $. The electron and hole states are respectively indicated in blue and red.} 
\label{fig1}
\end{figure}

The low-lying energy levels of single-particle states in TMDs QDs with the magnetic field dependence are shown in Fig. \ref{fig1}. The energies with and without magnetic field are calculated separately by using different formulas in Sec. \ref{subsec:nomag} and Sec. \ref{sec:mag}, respectively. One can see that the energy transition is smooth in the zero magnetic field limit, indicating that our analytical solutions for the eigenvalues with and without magnetic field match each other. 
This energy spectra are discrete, which is a direct consequence of finite-size confinement, comparing with the bulk energy bands. 
Unlike a conventional semiconductor QD \cite{tapsh1999129,tapsh2006energy}, the ground state of the conductance band is not  {degenerate} in zero magnetic field, two spin states have a gap of $\sim 0.1$ meV due to the spin-sublattice coupling. As the magnetic field increases, the two spins states are further separated a bit, the gap is about $0.6$ meV up to $30$ T. 
Eventually, Landau levels form with escalating magnetic field intensities. Notably, the spin splitting for hole states is much more significant, about $150$ meV for \ce{MoS2} and $430$ meV for \ce{WS2}. This motivated our decision to use electronic states with two spins and spin-up hole states as the basis for constructing the Hilbert space in ED.

\subsection{Interlayer Exciton Energy spectra}\label{sec:energy}

\begin{figure}[htp]
\centering
\includegraphics[width=8.4cm]{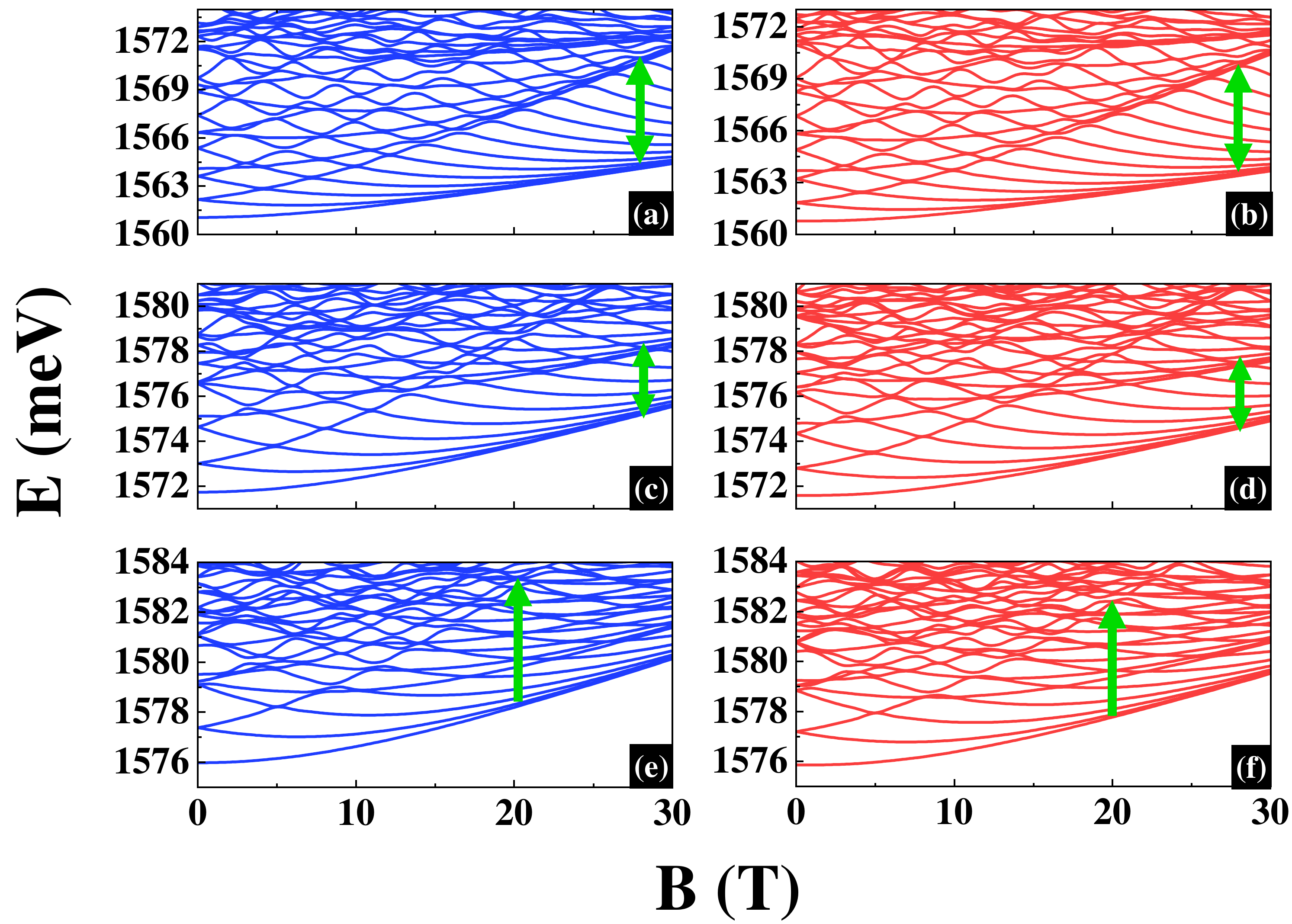}
\caption{The dependence of low-lying energy spectra of interlayer excitons on the magnetic field, where the electron and hole are confined within \ce{MoS2}/\ce{MoS2} homobilayer TMD QDs with a radius of $20$ nm. 
The left panels represent the excitons containing spin-up electron, while the right panels are for spin-down, indicating distinct spin arrangements. Panel (a) and (b) display the interlayer exciton energy spectra with a $5$ nm interlayer distance. Panels (c) and (d) indicate the exciton spectra at a $10$ nm interlayer distance, while panels (e) and (f) illustrate the exciton spectra at a $15$ nm distance.
}
\label{fig2}
\end{figure}

\begin{figure}[htp]
\centering
\includegraphics[width=8.4cm]{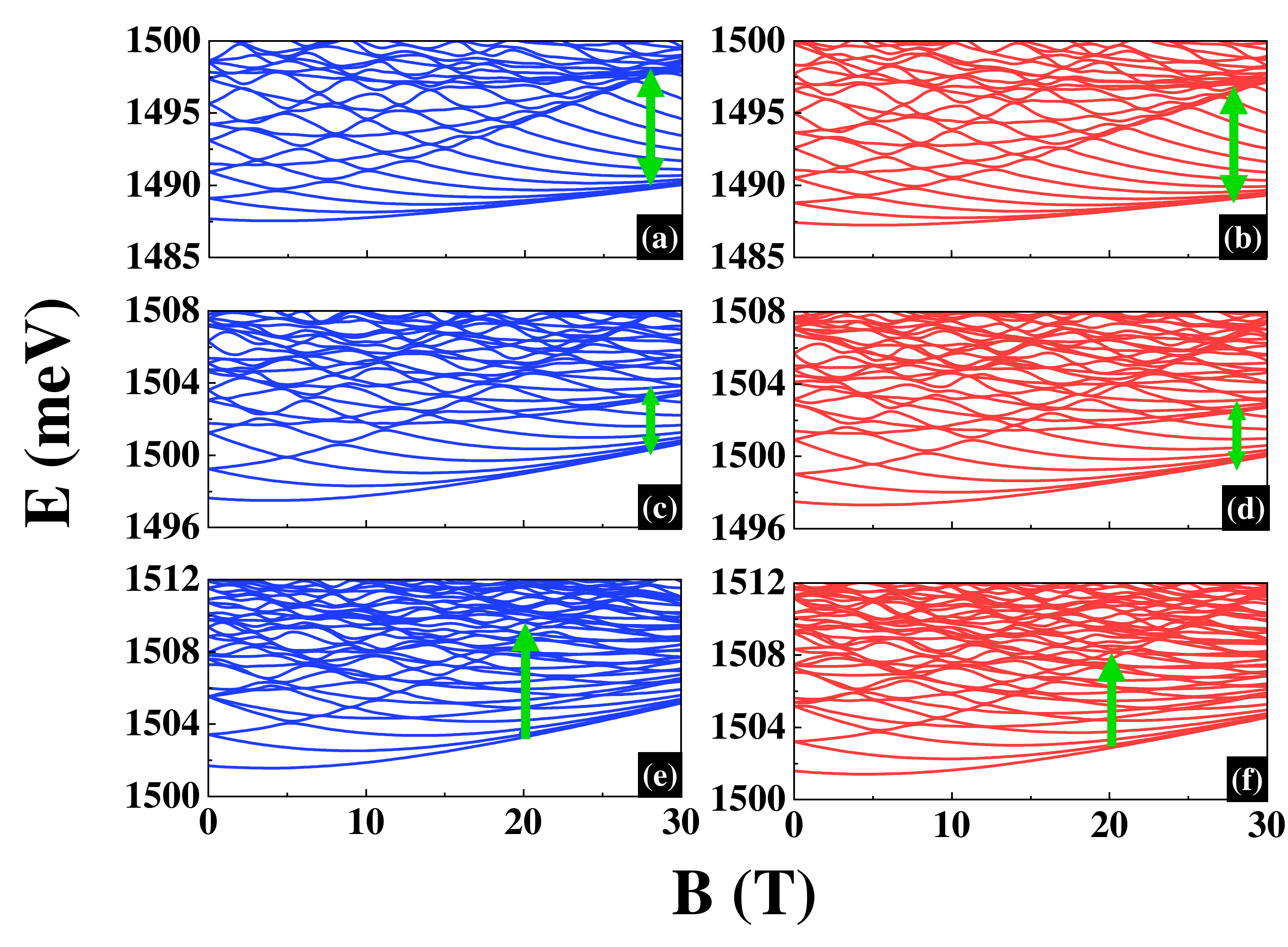}
\caption{Simialr to Fig. \ref{fig2}, the dependence of low-lying energy spectra of interlayer excitons on the magnetic field, where the electron and hole are confined within \ce{MoS2}/\ce{WS2} heterobilayer QDs with radius of $20$ nm, respectively. The left and right panels represent the excitons containing electrons with spin-up and spin-down, respectively. Panel (a) and (b) display the interlayer exciton energy spectra with a $5$ nm interlayer distance. Panels (b) and (c) for a $10$ nm interlayer distance, while panels (d) and (e) for a $15$ nm distance.}
\label{fig3}
\end{figure}

After characterizing the single-particle states in TMD QDs, we proceed to extend our investigation to two-particle systems, namely electron-hole pairs within double-layer QDs. For accurate numerical calculations of quantum systems affected by Coulomb interactions, it is crucial to choose an appropriate set of eigenstates as the basis. 
The radii of the QDs in electron and hole layers are the same, $R_{e} =R_h= 20$ nm. The radius is selected 
to manage the computational complexity. In Figs. \ref{fig2} and \ref{fig3}, we illustrate the dependency of interlayer exciton energy on the magnetic field with varying interlayer distances. 
In Fig. \ref{fig2}, we can clearly observe that Coulomb interactions reduce the energy of excitons, and this reduction diminishes with the increase in interlayer separation, since the Coulomb interaction is softened by the interlayer distance. Moreover, with the increase of the magnetic field, the single exciton still illustrates the Landau-type levels. It implies that quantum Hall effects may occur in these bosonic systems. In fact, the related phenomenon has been observed experimentally in bilayer TMDs system\cite{du2023fqhe}. The Landau level gaps in large magnetic fields are suppressed by the electron-hole Coulomb interaction, but will be recovered to the values of the non-interacting case when the separation of the two layers approaches to infinity. Notably, the Landau level gaps seem not increased monotonically with the increase of the distance $d$ between the two layers. As shown in Fig. \ref{fig1}, the first Landau level gap with $d=5$ nm is obviously larger than that with $d=10$ nm. However, when $d$ increases to $15$ nm, this gap increases and is more like the non-interacting case.

In Fig. \ref{fig3}, we extend our analysis to \ce{MoS2}/\ce{WS2} heterobilayer, following the same context as depicted in Fig. \ref{fig2}. These results collectively illustrate the dependency of interlayer exciton energy on the magnetic field across distinct material systems. 
We observe that the energy of the low-lying excitonic states formed in the heterobilayer is lower than that in the homobilayer. This is attributed to the band mismatch between the conduction and valence bands while different materials are stacked. It leads to the formation of a type-II band alignment\cite{bandtype} where the hole is confined in \ce{WS2} QDs with lower single particle energy. 
The Landau-type levels behavior of excitons in magnetic field exhibits characteristics similar to that in the homobilayer QDs.

\begin{figure}[htp]
\centering
\includegraphics[width=8.4cm]{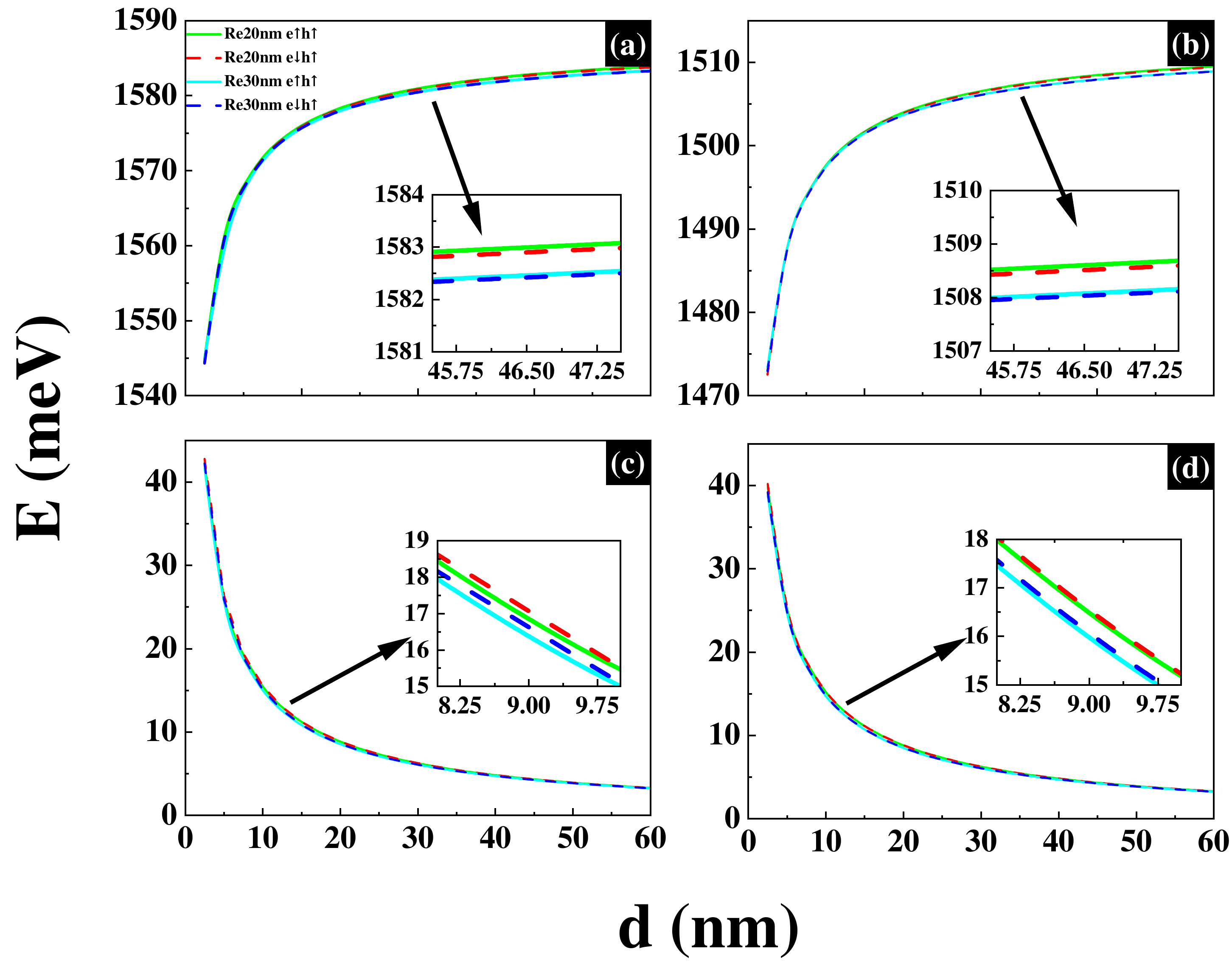}
\caption{The ground-state and binding energies of the interlayer excitons vary with the interlayer distance in the absence of a magnetic field: (a) Ground state energy of \ce{MoS2}/\ce{MoS2} homobilayer QDs; (b) Ground state energy \ce{MoS2}/\ce{WS2} of heterobilayer QDs; (c) Binding energy of \ce{MoS2}/\ce{MoS2} homobilayer QDs; (d) Binding energy \ce{MoS2}/\ce{WS2} of heterobilayer QDs. The solid line represents the energy for an exciton with an upward-spin electron, while the dashed line corresponds to a downward-spin electron. Both equal and unequal radius configurations are illustrated. The insert panels provide a detailed zoom-in on the energies.}
\label{fig4}
\end{figure}

The ground state energy of interlayer exciton as a function of distance is shown in Fig. \ref{fig4} for \ce{MoS2}/\ce{MoS2} homobilayer and \ce{MoS2}/\ce{WS2} heterobilayer QDs. 
As shown in Figs. \ref{fig4} (a) and (b), it is apparent that the excitonic energy profile contains an initial augmentation followed by a slowly varied function of the interlayer separation. 
Consequently, from Figs. \ref{fig4} (c) and (d), 
a stronger bound energy at smaller interlayer distances. 
Nevertheless, the long-range nature of Coulomb forces \cite{fu2024bilayer} remain influential at a larger distance. 
In comparison, we also show the results that the QDs radii are inconsistent, where the electron is in a $30$-nm-radius dot and the hole locates in a $20$-nm-radius dot. 
Both the exciton ground-state energy and binding energy slightly decrease compared to QDs with uniform radii. This reduction occurs because electron confined in larger QDs. 

\begin{figure}[htp]
\centering
\includegraphics[width=8.40cm]{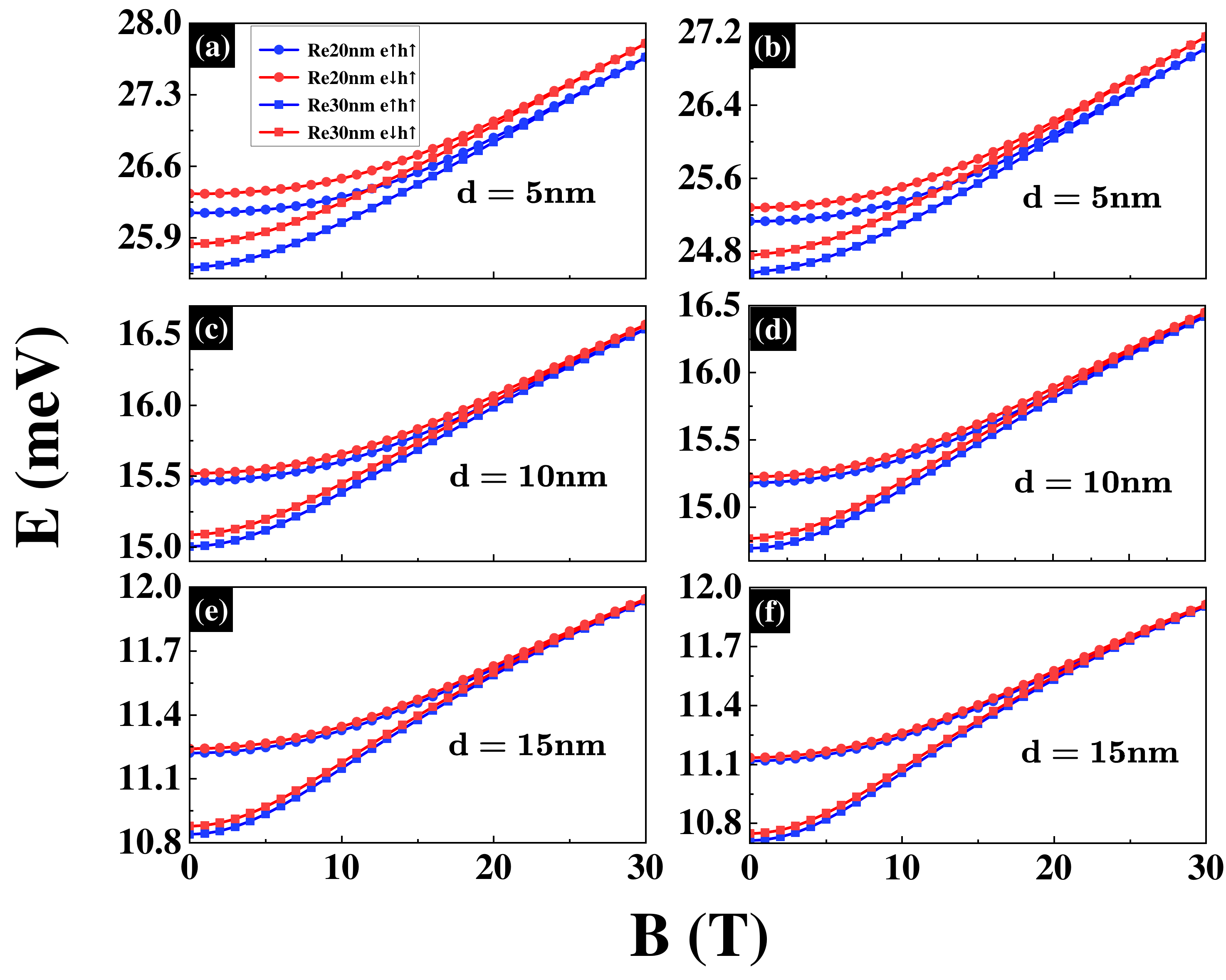}
\caption{Magnetic field dependence of binding energy of interlayer exciton ground state. The left panels are for \ce{MoS2}/\ce{MoS2} homobilayer QDs, while the right panels shows results for \ce{MoS2}/\ce{WS2} heterobilayer QDs. The interlayer distance are (a) and (b) $ d = 5$ nm, (c) and (d) $ d = 10$ nm, (e) and (f) $ d = 15$ nm. Different electron spin configurations in the excitons are indicated by distinct colors, blue and red denote the exciton with electron spin up and down respectively. Circle symbol represents that two QDs have the same radius $R_e=R_h=20$ nm, while the square symbol indicates $R_e=30, R_h=20$ nm. }
\label{fig5}
\end{figure}

More results are presented in Fig. \ref{fig5}, showing the exciton binding energy as a function of the magnetic field. The binding energies all increase with the magnetic fields, which is attributed to the increase of the cyclotron motion, and decrease as the interlayer distance grows. 
Another notable feature is that excitons with spin-up electrons have lower binding energies than those with spin-down electrons. It is attributed to the different wave functions for different spins induced by the spin-valley coupling.

We also investigate a double-layer QDs stacked with different radii, the upper QD providing an electron with radius $30$ nm and the lower QD containing a hole with radius $20$ nm. In this case, as shown in Fig. \ref{fig5}, the binding energies in weak magnetic fields are much lower than those in the double-layer QDs with identical radius $20$ nm. The single-particle density of states in larger QDs is higher, and the non-interacting energy decreases more significantly than the Coulomb interaction, although the Coulomb interaction is also decreased by involving more single-particle states. When the magnetic field is strong, the single-particle states approach to be degenerate in Landau levels, causing the binding energies for QDs of different sizes to converge.

\subsection{Electron hole separation}\label{sec:seperation}
The Coulomb interaction spatially constrains the electrons and holes, differing from intralayer excitons.
In the $xOy$ plane, the electron-hole separation is expressed as 
\begin{equation}
\left\langle \mathbf{r}_{e}\mathbf{-r}_{h}\right\rangle=\sum_{i,k}C_{i}^{\ast }C_{k}\left\langle i_{e}; i_{h}\right\vert \mathbf{r}_{e}\mathbf{-r}_{h}\left\vert k_{e};k_{h}\right\rangle.
\end{equation} 
Here, $C_{i}$ and $C_{k}$ represent the expansion coefficients of the ground state of the exciton obtained by diagonalizing the Hamiltonian, $|GS\rangle = \sum_i C_{i} |i_{e};i_{h}\rangle$.
This measure provides insight into the spatial distribution and binding nature of the exciton. 

\begin{figure}[htp]
\centering
\includegraphics[width=8.60cm]{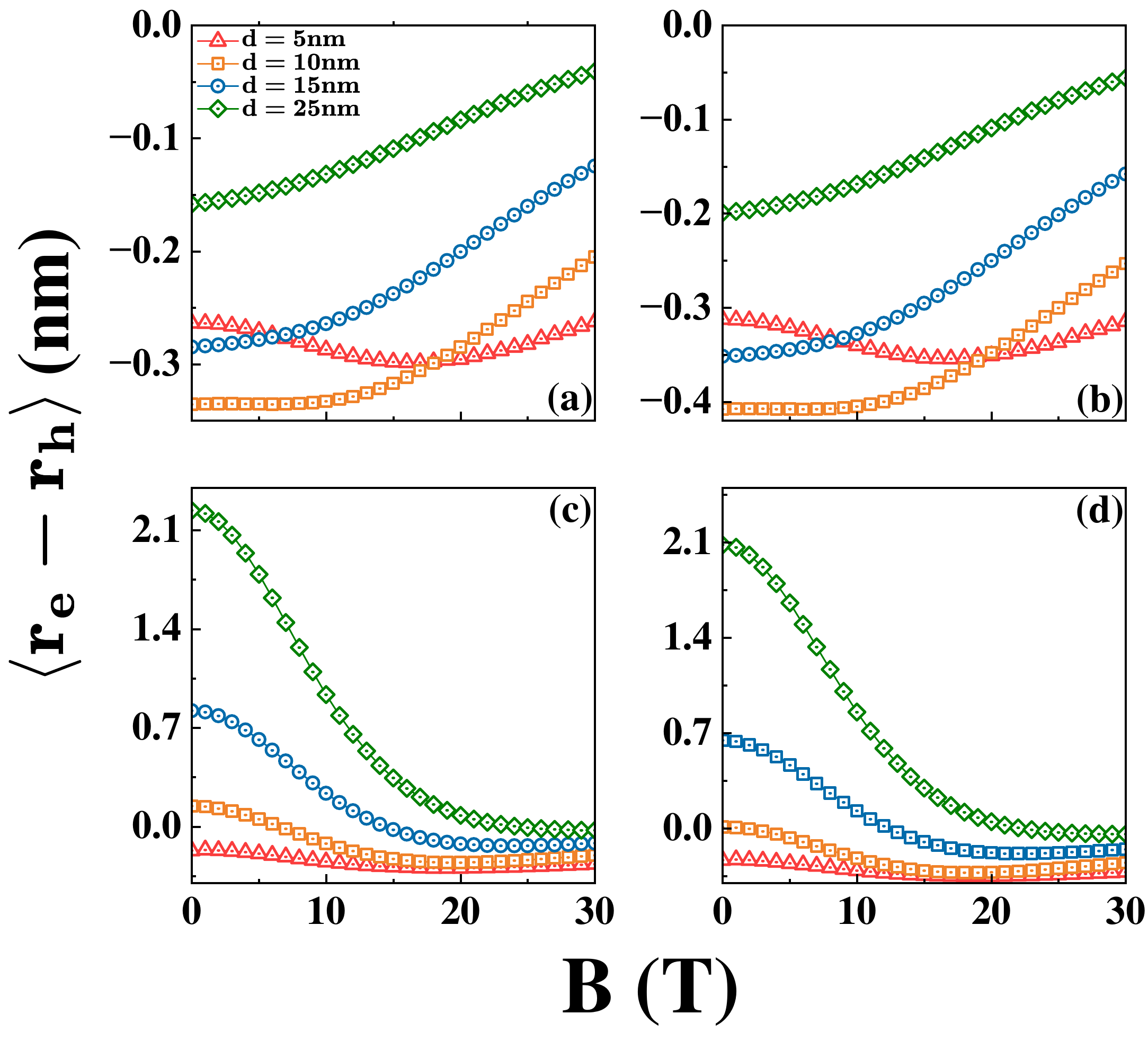}
\caption{Evolution of the ground-state electron-hole separation $ \left\langle \mathbf{r}_{e}\mathbf{-r}_{h}\right\rangle $ with magnetic fields in different distances for \ce{MoS2}/\ce{WS2} homobilayer. The top panels show separations in QDs with identical $20$ nm radius, while the bottom panels show separations in QDs with $R_{e} = 30$ nm in electron layer and $R_{h} = 20$ nm in hole layer. The interlayer exciton separation for different electron spin: (a) and (c) electron spin-up, (b) and (d) electron spin-down.}
\label{fig6}
\end{figure}

In Fig. \ref{fig6}, we show the electron-hole separation as a function of magnetic fields in the \ce{MoS2}/\ce{WS2} heterobilayer QDs only. 
For homobilayer case, the electron is generally located exactly upon the hole, and the separation $\left\langle \mathbf{r}_{e}\mathbf{-r}_{h}\right\rangle$ approaches zero (up to $\sim 0.06$ nm).
Comparing Fig. \ref{fig6} (a) with (b) and (c) with (d), the separation of excitons with different spin electrons is minimal. Fig. \ref{fig6} (a) and (b) show results with identical radius $20$ nm of the electron and hole layers, indicating that the separation magnitude is much smaller than the system’s size. 
In contrast, as shown in Figs. \ref{fig6}(c) and (d), the separation of the system with $R_e=30$ nm and $R_h=20$ nm varies significantly with magnetic field. In weak magnetic field, the separation is one magnitude larger than the case of identical radius, since the orbit of low-energy single-particle wave function are different. In large magnetic field, due to the Landau quantization, the difference of separation between the two cases is not obvious.

Notably, unlike the energy of the ground state, the absolute value of separation $|\left\langle \mathbf{r}_{e}\mathbf{-r}_{h}\right\rangle |$ does not vary monotonously with either the magnetic field or the distance between the two layers. With increase of distance, the separation starts to increase and then decreases, in the case of identical radius of two QDs. However, the situation is reversed in the case of different radii of two QDs. As the magnetic field varies, the change of the $|\left\langle \mathbf{r}_{e}\mathbf{-r}_{h}\right\rangle |$ becomes even more complicated. Comparing to the excitons in conventional semiconductor QDs, the anomalous change of electron-hole separation as well as the density shape of exciton here is attributed to the complex orbit of the wave function induced by SOC in the system, which may be observed by detecting the wave functions orbitals \cite{camenzind2019spectroscopy}.

%

\subsection{Pseudospin textures of interlayer excitons}\label{sec:topo}

Topological textures of spin fields in QDs are induced by SOC which also plays an important role in TMDs. 
Studying topological features in QDs boosts spintronics by refining spin control and enhances quantum information processing by improving qubit storage and manipulation \cite{luo2024encyclopedia}. 
Similar to spin, the two sublattices of TMDs can be represented by pseudospin.
In the Hamiltonian of TMDs QDs, the pseudospin-orbit coupling, similar to Rashba and Dresselhaus SOC, also results in a non-trivial pseudospin texture. 
The pseudospin fields for electron and hole in an exciton can be defined as follows
\begin{eqnarray}
\sigma _{x,y}^{\alpha }\left( \mathbf{r}\right)  &=& \sum_{i,k}C_{i}^{\ast
}C_{k}\delta _{i\alpha,k\alpha }\left( \chi _{2i}^{\alpha \ast }\chi
_{1k}^{\alpha }\pm \chi _{1i}^{\alpha \ast }\chi _{2k}^{\alpha }\right)   \notag \\
&\times& \exp \left( i\theta \left( \pm \tau +j_{k_{\alpha }}-j_{i_{\alpha
}}\right) \right), 
\end{eqnarray}
and the density is given by 
\begin{eqnarray}
n^{\alpha }\left( \mathbf{r}\right)  &=& \sum_{i,k}C_{i}^{\ast }C_{k}\delta
_{i\alpha ,k\alpha }\left(  \chi _{1i}^{\alpha \ast }\chi _{1k}^{\alpha }+\chi _{2i}^{\alpha \ast }\chi
_{2k}^{\alpha } \right)  \notag \\
&\times& \exp \left( i\left( j_{k_{\alpha }}-j_{i_{\alpha }}\right) \theta \right),
\end{eqnarray}
where $ \alpha $ denote the electron or hole and $ \phi^{a} $ correspond to two different spinors. The in-plane pseudospin field is represented by $ \mathbf{\sigma^{\alpha}(r_{\alpha})} = (\sigma^{\alpha}_{x},\sigma^{\alpha}_{y}) $.

\begin{figure}[htp]
\centering
\includegraphics[width=8.40cm]{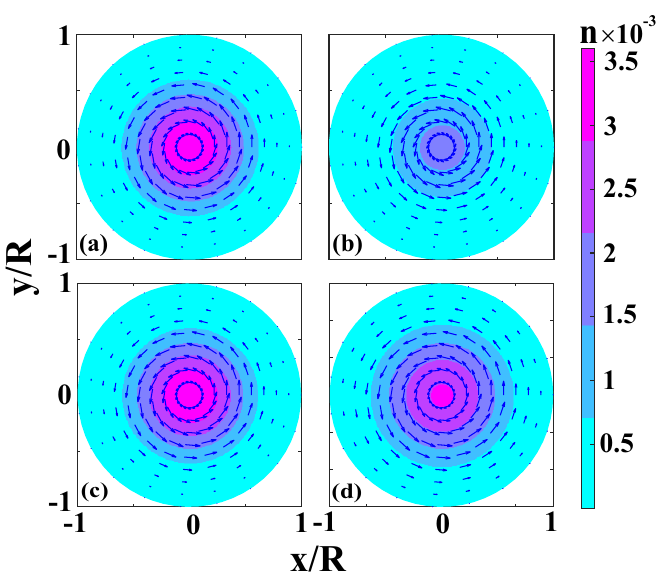}
\caption{Ground state pseudo-spin fields $(\sigma^{\alpha}_{x},\sigma^{\alpha}_{y}) $ of interlayer excitons with electron and hole spin-up at magnetic field $ B = 1$ T. (a) and (c) The pseudospin textures of the electron and the hole in interlayer exciton in \ce{MoS2}/\ce{MoS2} homobilayer TMDs QDs with radii $ R_{e} = R_{h}= 20\text{nm}$, respectively. The distance between the two layers is $5$ nm. (b) and (d) The pseudospin textures of the electron and the hole in interlayer exciton in \ce{MoS2}/\ce{WS2} heterobilayer QDs with $ R_{h} = 20\text{nm} $ and  $R_{e} = 30\text{nm} $. The distance between the two layers is $10$ nm.}
\label{fig7}
\end{figure}

\begin{figure}[htp]
\centering
\includegraphics[width=8.40cm]{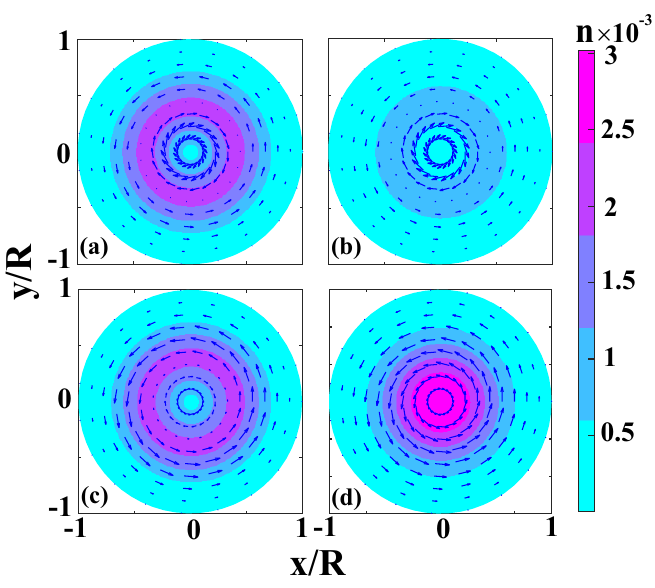}
\caption{The same as Fig. \ref{fig7}, but depicting the pseudospin texture for the first excited state under a magnetic field of $B = 5\text{T}$.}
\label{fig8}
\end{figure}

In Fig. \ref{fig7}, two examples of the pseudospin fields of the ground states are displayed. The pseudospins of electron and hole are textured with nontrivial topological charge $1$ \cite{luo2019tuning}. The topological spin fields in QDs induced by Rashba or Dresselhaus SOC usually have zero vorticity, while both topological charge and vorticity are nonzero here, since exciton is in $K$ valley with $\tau=1$. We note that in $K'$ valley, the pseudospin textures are more like those with Dresselhaus SOC given that $\tau=-1$. The pseudospin textures of the first excited state of the exciton are shown in Fig. \ref{fig8}. Similar as the QD with Rashba or Dresselhaus SOC, the topological charge of pseudospin remains unchanged but its vorticity is inverse. 

\subsection{Light absorption}\label{sec:light} 

Far-infrared (FIR) magneto-optical absorption spectroscopy is a technique used to study and utilize the optical properties of materials. The radius of the QD discussed here is typically only tens of nanometers, significantly smaller than the wavelength of far-infrared light corresponding to the energy gaps of the system. Therefore, we can compute exciton light absorption using the dipole approximation \cite{avetisyan2012strong,avetisyan2013superintense,chakraborty2018controllable}. 
The incident light is supposed to be perpendicular to the plane of TMDs. The dipole transition matrix element from state $| k\rangle$ to state $|i \rangle$  denoted as $D_{ik}=\left\langle i\left\vert \bm{\varepsilon }\cdot \bm{r}\right\vert	k\right\rangle $, where $\bm{\varepsilon }$ represents the polarization vector lying in the $xOy$ plane. The dipole transition matrix element depends on the polarization of the incident light. Generally, we consider the incident light to be unpolarized. For a single particle in QDs, the dipole transition matrix element is given by 
\begin{eqnarray}
D_{ik} &=&
\pi (\Omega _{x} + i\Omega_y) \int r^{2}dr\left( \chi _{1i}^{\ast }\chi_{1k}+\chi _{2i}^{\ast }\chi _{2i}\right), 
\end{eqnarray}
where $\Omega_{x,y}= \delta _{m_{k}+1,m_{i}}\pm \delta_{m_{k}-1,m_{i}} $, $m$ is the orbital angular momentum quantum number and $ \chi _{12} $ is spinor corresponds to two sublattices. 

According to the transition selection rule, only transitions with $\Delta m=m_{i}-m_{k}=\pm 1$ are allowed between initial state $\left\vert i\right\rangle $ and final state $\left\vert k\right\rangle $. We calculate the amplitude of the dipole transition matrix element as $A_{k\rightarrow i}=\sum_{i,k}C_{i}^{\ast }C_{k}\left\langle i_{e},i_{h}\left\vert \mathbf{r}_{h}-\mathbf{r}_{e} \right\vert k_{e},k_{h}\right\rangle .$ The intensity of absorption is proportional to $\left\vert A_{k\rightarrow i}\right\vert ^{2}$. We consider only the transition from ground state to excited state.

\begin{figure}[htp]
\centering
\includegraphics[width=8.40cm]{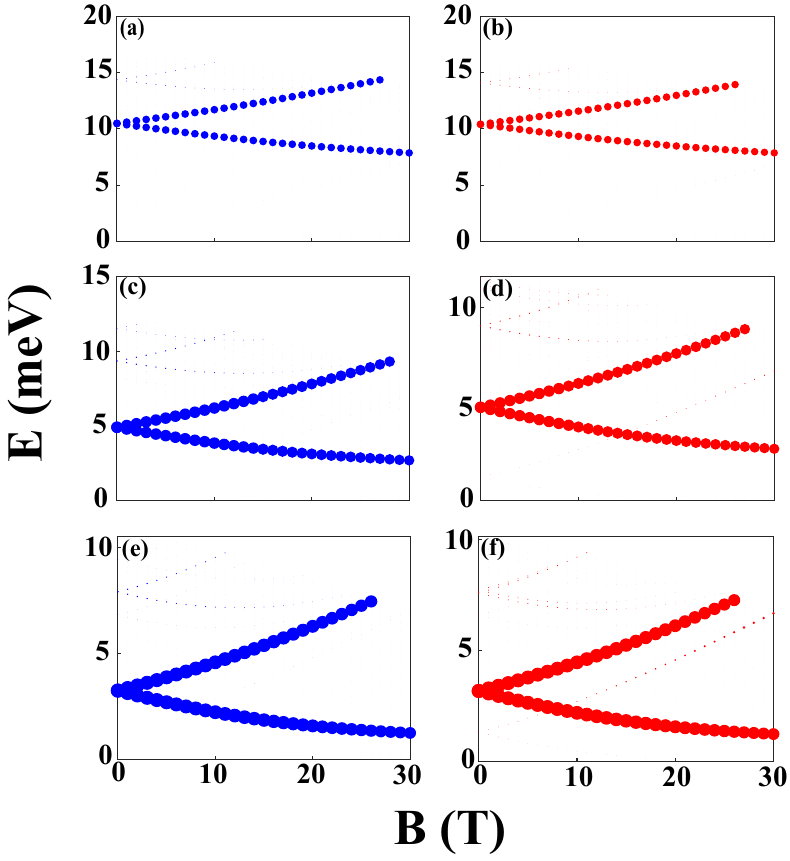}
\caption{Dipole-allowed optical absorption spectra of \ce{MoS2}/\ce{MoS2} homobilayer QDs for various interlayer distances. The left panels show the results with electron spin-up, while the right panels illustrate for spin-down. (a) and (b) display results with interlayer distance of $ d = 5$ nm, (c) and (d) for an interlayer distance of $ d = 10$ nm, and (e) and (f) for $ d = 15$ nm. The radii of the QDs in both layers are $20$ nm. The size of the scatterers is proportional to the intensity of the light absorption.}
\label{fig9}
\end{figure}

\begin{figure}[htp]
\centering
\includegraphics[width=8.40cm]{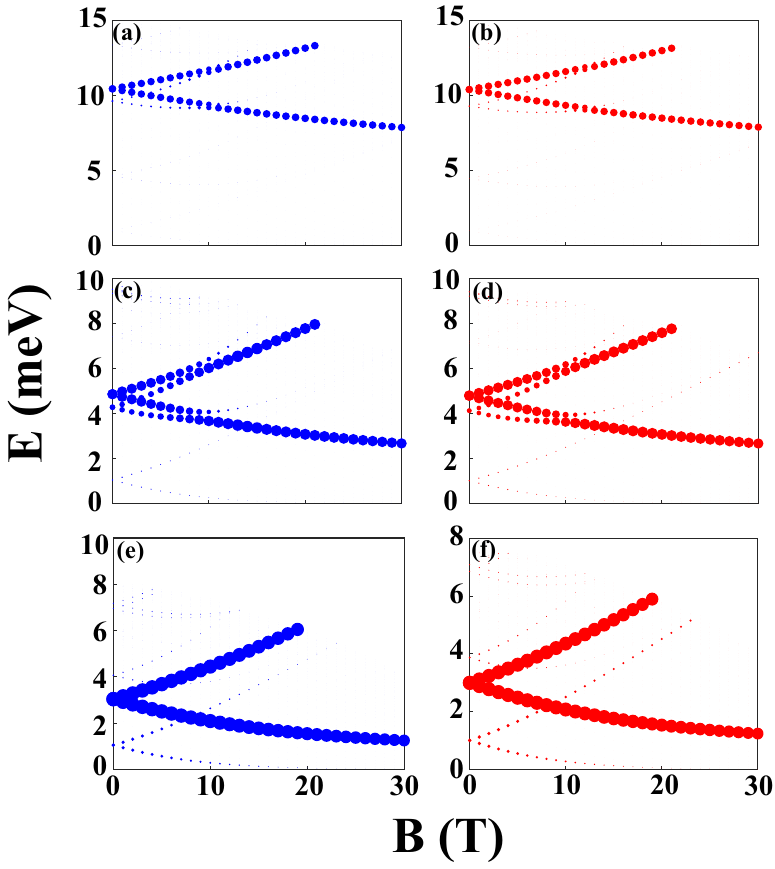}
\caption{The same dipole-allowed optical absorption spectra as Fig \ref{fig9}, but the radii of electron and hole layer are $ R_{e} = 30$ nm and $ R_{h} = 20$ nm, respectively.}
\label{fig10}
\end{figure}

Fig. \ref{fig9} illustrates the dipole-allowed light absorption of interlayer excitons across different interlayer distances under varying magnetic fields. 
Without interactions, some transition modes are allowed by the selection rules, but Coulomb interactions make them forbidden. Comparing cases with different interlayer distances, we notice that dipole-allowed transitions in systems with smaller interlayer distances  favor modes with higher energy. The intensity of transition modes increases with larger interlayer distances. Moreover, Fig. \ref{fig10} shows dipole-allowed optical absorption spectra of \ce{MoS2}/\ce{MoS2} homobilayers for QDs with radii $ R_{e} = 30\text{nm} $ and $ R_{h} = 20\text{nm} $. We find that light absorption varies with interlayer distance similarly to Fig. \ref{fig9}, but more transition modes available.

\section{Results on biexciton} \label{sec:biexciton}

We have discussed the physical quantities of a single interlayer exciton in the previous section. 
It is worthy exploring the case of interacting interlayer excitons, which is described by the full many-body Hamiltonian in Eq.\eqref{many-bodyh}. For simplicity, we here consider a  
biexciton system 
containing two electrons and two holes. The electrons and holes still locate in the upper and lower layers, respectively.
The Hamiltonian contains intra-layer electron-electron and hole-hole interactions comparing with that of a single exciton system. We discuss two scenarios, where electrons either have identical or opposite spins and the spin of holes is fixed. 
The biexciton system exhibits more complex interactions than the single exciton system, resulting in different features.

\begin{figure}[htp]
\centering
\includegraphics[width=8.40cm]{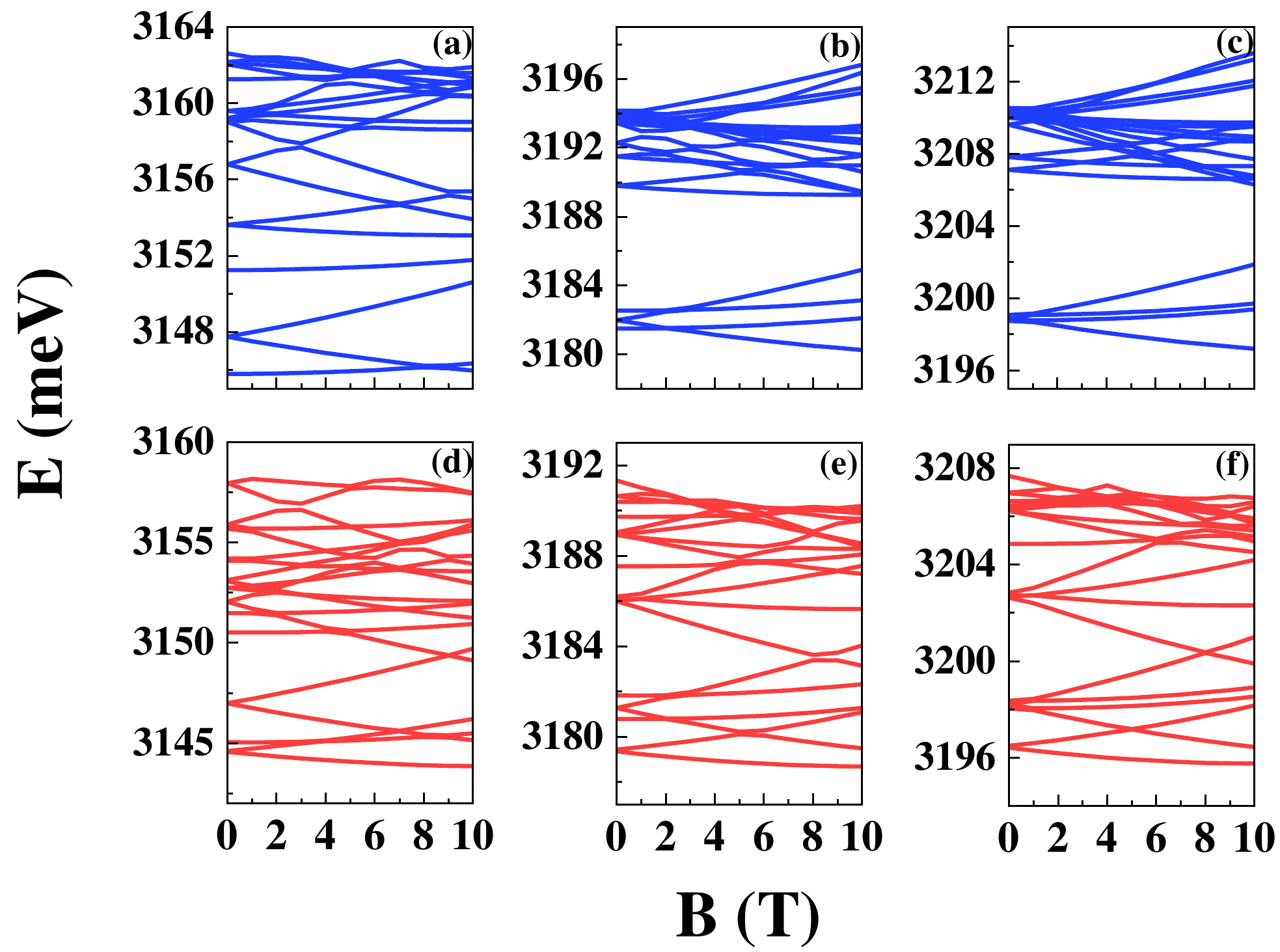}
\caption{The dependence of low-lying energy spectra of biexcitons on the magnetic field, where the electrons and holes are confined within \ce{MoS2}/\ce{MoS2} homobilayer QDs with the same radius of $10$ nm. (a) to (c) represent the system containing two electrons with the same spin, while the (d) to (f) are for two electrons with different spins. Panels (a), (b) and (c) display the energy spectra with interlayer distances $5$ nm, $10$ nm and $15$ nm respectively. Panels (d), (e) and (f) are for the same distance as (a) to (c), respectively.}
\label{fig11}
\end{figure}

\begin{figure}[htp]
\centering
\includegraphics[width=8.40cm]{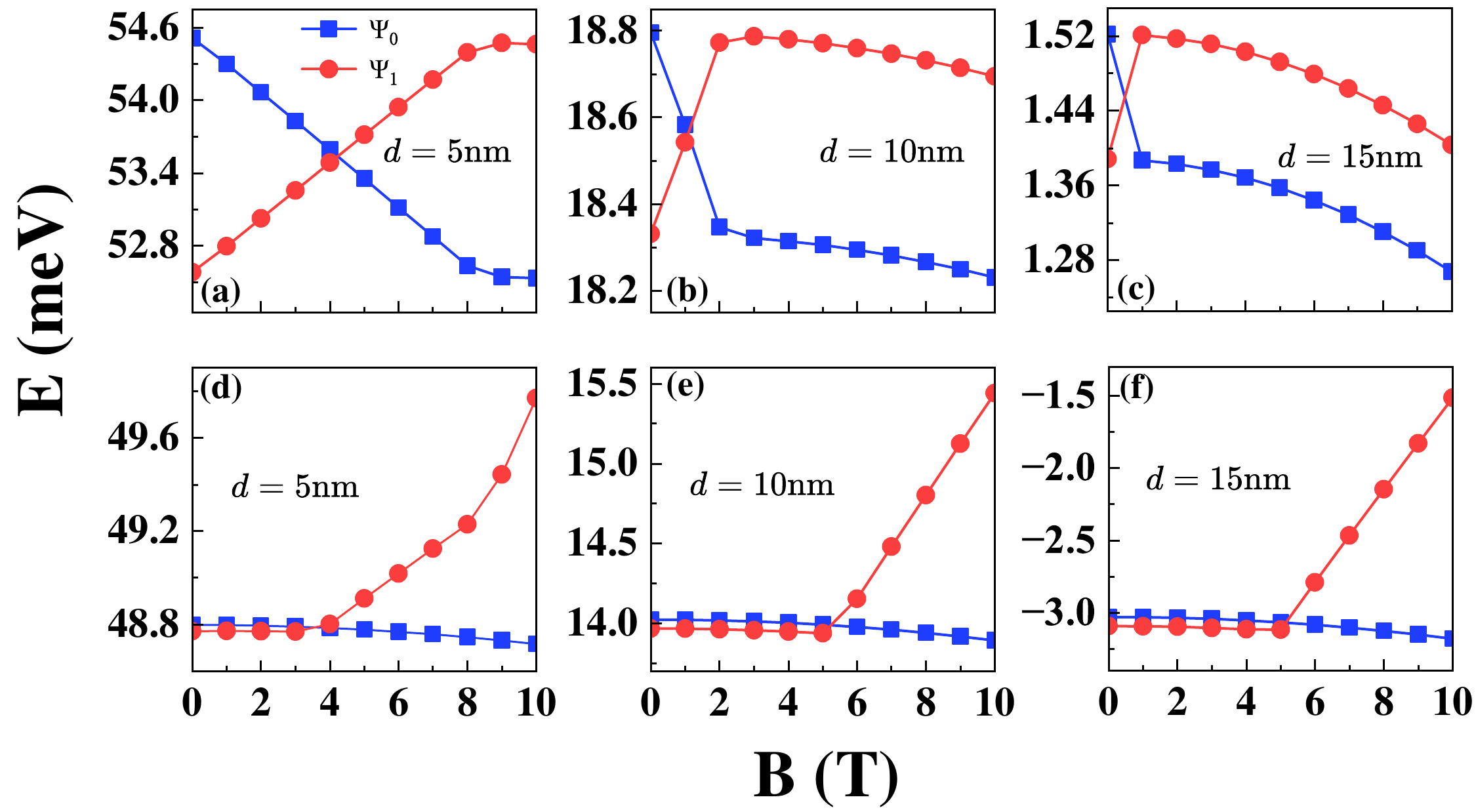}
\caption{Magnetic field dependence of binding energy of biexciton ground state and the first excited state. The panels (a) to (f) correspond to the cases in Figs. \ref{fig11}(a) to (f), respectively. (a) to (c) represent the biexciton containing two electrons with spin-up, while (d) to (f) are for two electrons with different spins.}
\label{fig12}
\end{figure}

Again, by employing the ED, the energy spectra are shown in Fig. \ref{fig11}, indicating rich energy-level structures due to many-body correlations. As the interlayer distance increases, the ground state and low-lying excited states exhibit higher energy. 
Remarkably, the four low-lying states are well gapped from higher states with sufficiently large interlayer distance when the two electrons have the same spin, as shown in Figs. \ref{fig11}(b) and (c). By comparing the energy spectra in Figs. \ref{fig11} (a) to (c) and Figs. \ref{fig11}(d) to (f), when the electrons have different spins, we find that some levels emerge. Particularly, the new ground state with two-fold degeneracy at zero magnetic field due to the Kramers pair appears, and its energy is suppressed significantly by the exchange interaction between two different spins. Moreover, the gap between low-lying levels and higher levels become not obvious, it only exists in weak magnetic fields, as shown in Figs. \ref{fig11}(e) and (f).
For more information, the binding energies of the ground state and the first excited state are displayed in Fig. \ref{fig12}. 
The mutations in binding energies arises from the transition of ground state with increase of magnetic field. 

\begin{figure}[htp]
\centering
\includegraphics[width=8.40cm]{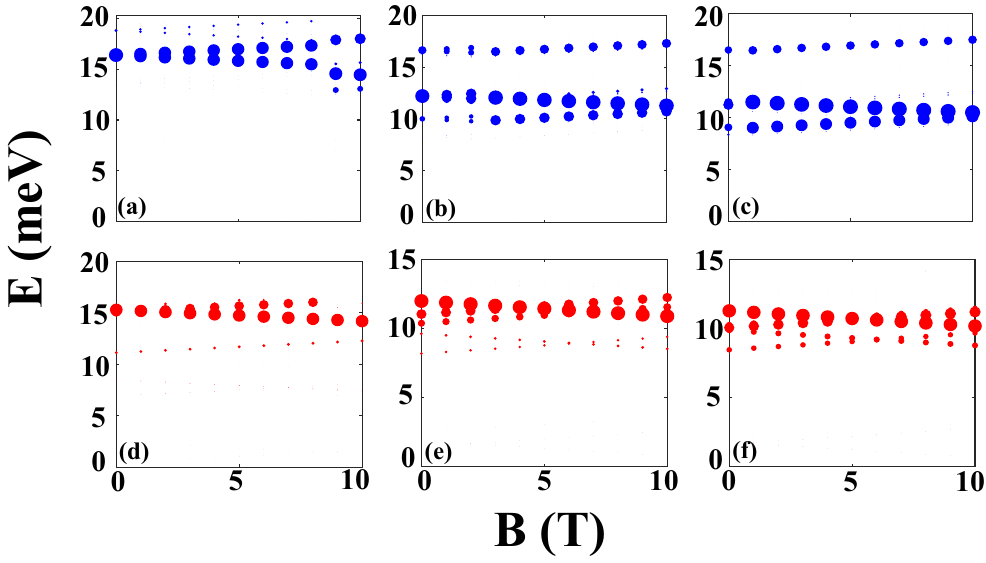}
\caption{Dipole-allowed optical absorption spectra of biexciton for various interlayer distances. (a) to (c) represent the biexciton containing two electrons with spin up, while (e) to (f) are for two electrons with different spin. Panels (a) and (b) display results with interlayer distance of $ d = 5$ nm, (c) and (d) for an interlayer distance of $ d = 10$ nm, and (e) and (f) for $ d = 10$ nm. The radii of the QDs in both layers are $ 15$ nm.}
\label{fig13}
\end{figure}

The dipole-allowed light absorptions of biexciton systems across different interlayer distances under varying magnetic fields are shown in Fig. \ref{fig13}. Comparing with the case of single interlayer exciton, more absorption modes are visible. The mutation of the absorption intensity is  attributed to the transition of the ground state when the two electrons have the same spin. For the case that two electrons have different spins, no transition occurs in ground state. The level crossing in higher levels does not significantly influence the absorption. Consequently, by measuring the light absorption spectra, one can distinguish the electrons spins in the biexciton ssytems.

It is noted that only biexcitons in homobilayer QDs are studied here. The energy spectra and light absorptions should remain qualitatively unchanged in heterobilayer QDs, as the physical properties of single exciton (with the exception of the electron-hole separation) are not significantly altered by the transition from homobilayer to heterobilayer QDs.

\section{SUMMARY} \label{sec:summary}
We constructed a double-layer TMDs QDs model with adjustable interlayer distance based on analyzing a single particle in hard wall potential TMDs QD subjected to an external magnetic field. This nanostructure enables the formation of interlayer excitons, which can be achieved through the modulation of band energies via external voltage or by leveraging the intrinsic properties of TMDs, represent one or more interlayer excitons with electron-electron, hole-hole and electron-hole correlations. By employing ED, we numerically explore this interlayer system with Coulomb interaction.
We report on several characteristics of an interlayer exciton, including energy spectrum, binding energy, spatial distribution of electron and hole topological pseudospin textures, and optical absorption spectrum with external magnetic fields in adjustable interlayer distances. 
Remarkably, the Landau level gap, the electron-hole separation in an exciton varies non-monotonously with the increase of the interlayer distance, due to the sublattice pseudospin-orbit coupling in TMDs. Moreover, the ground state and excitation states of the interlayer exciton are all pseudospin textured topologically by this pseudospin-orbit coupling. 

Furthermore, we studied many-body systems with two interacting interlayer excitons which contains two electrons and two holes. The energy spectra and light absorptions are also numerically calculated. 
Our numerical calculations show unique energy spectra with interaction-driven spectral splitting and enhanced binding energies. The optical absorption shifts under magnetic fields emerge, which are not observed in simpler systems. These findings highlight the tunable potential of TMD QDs for advanced quantum and optoelectronic technologies, allowing precise control of excitonic properties through external fields and structural tuning. 
Our detailed analysis of their response to magnetic fields and interlayer distance fills in gaps in knowledge about the mechanism of formation and possible consequences within the nanoelectronics and quantum computing research domains.

\begin{acknowledgements}
This work was supported by the NSF-China under Grant No. 11804396. The authors acknowledge Chuanchun Shu, Yutao Hu, and Yi Jiang for helpful discussions. We are grateful to the High Performance Computing Center of Central South University for partial support of this work.
\end{acknowledgements}

\appendix

\section{Band structure} \label{app:band}

The bands selection scenario is illustrated schematically in Fig. \ref{band}. The valley is supposed to be polarized. Electrons and holes are from different layers. 

\begin{widetext}

\begin{figure}[htp]
\centering
\includegraphics[width=8.40cm]{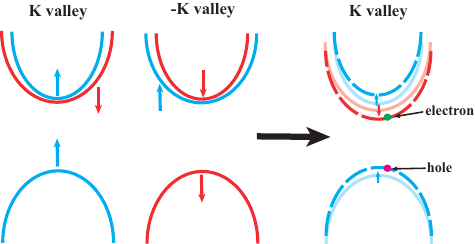}
\caption{The schematic bands structure of TMDs near Fermi surface. The lower bands in valence band are neglected since they are far away from the Fermi surface. Only one valley, say valley $K$, is considered. The right panel indicates the bands shift by Zeeman coupling in a magnetic field. Electrons and holes are marked in different bands, meanwhile in different layers. }
\label{band}
\end{figure}

\section{Derivation of basis} \label{app:basis}
In this appendix, we derive the eigenstates of TMDs QDs, which are utilized as the basis for deriving the matrix elements in our numerical calculations. The derivation begins with the general form of the TMDs QDs Hamiltonian, as detailed in Eq.~\eqref{eq:ham}
\begin{equation}
H=\left( 
\begin{array}{cc}
\frac{\Delta }{2} & \frac{at}{\hbar }\left( \tau \left( p_{x}+eA_{x}\right)
-i\left( p_{y}+eA_{y}\right) \right)  \\ 
\frac{at}{\hbar }\left( \tau \left( p_{x}+eA_{x}\right) +i\left(
p_{y}+eA_{y}\right) \right)  & -\frac{\Delta }{2}+\lambda \tau s%
\end{array}%
\right) +V\left( r\right) \left( 
\begin{array}{cc}
1 & 0 \\ 
0 & -1%
\end{array}%
\right) .	
\end{equation}
By using the the symmetry gauge $\textbf{A} = \frac{B}{2}\left(-y,x,0\right) $, we can get
\begin{subequations}
\begin{align}
\frac{at}{\hbar }\left( \tau \left( p_{x}+eA_{x}\right) -i\left(
p_{y}+eA_{y}\right) \right) =-iat\exp \left( -i\tau
\theta \right) \left( \tau \frac{\partial }{\partial r}-i\frac{1}{r}\frac{%
\partial }{\partial \theta }+e\frac{B}{2\hbar }r\right)  \\
\frac{at}{\hbar }\left( \tau \left( p_{x}+eA_{x}\right) +i\left(
p_{y}+eA_{y}\right) \right)  =-iat\exp \left( i\tau
\theta \right) \left( \tau \frac{\partial }{\partial r}+i\frac{1}{r}\frac{%
\partial }{\partial \theta }-e\frac{B}{2\hbar }r\right), 
\end{align}
\label{eq:polor_co}
\end{subequations}
then the Hamiltonian of TMDs QDs in the polar coordinates become
\begin{equation}
H=\left( 
\begin{array}{cc}
\frac{\Delta }{2} & -iat\exp \left( -i\tau \theta
\right) \left( \tau \frac{\partial }{\partial r}-i\frac{1}{r}\frac{\partial 
}{\partial \theta }+e\frac{B}{2\hbar }r\right)  \\ 
-iat\exp \left( i\tau \theta \right) \left( \tau \frac{%
\partial }{\partial r}+i\frac{1}{r}\frac{\partial }{\partial \theta }-e\frac{%
B}{2\hbar }r\right)  & -\frac{\Delta }{2}+\lambda \tau s%
\end{array}%
\right) +V\left( r\right) \sigma _{z}.
\end{equation}

\subsection{Zero magnetic field}
In the absence of a magnetic field, the Dirac equation can be reformulated into two coupled differential equations using the eigenstates defined in Eq.~\eqref{eq:waf} and expressed in dimensionless units.
\begin{subequations}
\begin{align}
at\left( \tau \frac{\partial }{\partial r}+\frac{1}{r}%
\left( j+\frac{\tau }{2}\right) \right) \chi_{2}(r)  &=\left( E-\frac{%
\Delta }{2}\right) \chi_{1}(r)  \label{eq:appsub1}\\
at\left( \tau \frac{\partial }{\partial r}-\frac{1}{r}%
\left( j-\frac{\tau }{2}\right) \right) \chi_{1}(r)  &=-\left( E+\frac{%
\Delta }{2}-\lambda \tau s\right)\chi_{2}(r) .
\end{align}
\end{subequations}	
We can deduce that  $ \frac{\left( \tau \frac{\partial }{\partial \rho }+\frac{1}{\rho }\left( j+%
\frac{\tau }{2}\right) \right)\chi_{2}(r) }{\left( \varepsilon -\frac{1}{2}\delta \right) }=\chi_{1}(r) $
and subsequently apply it to Eq.~\eqref{eq:appsub1}, and $ \chi _{2}\left( \rho \right) $ yields the following differential equation 
\begin{equation}
\rho ^{2}\frac{\partial }{\partial \rho ^{2}}\chi _{2}\left( \rho \right)
+\rho \frac{\partial }{\partial \rho }\chi _{2}\left( \rho \right) -\left( j+
\frac{\tau }{2}\right) ^{2}\chi _{2}\left( \rho \right) +\kappa ^{2}\rho
^{2}\chi _{2}\left( \rho \right) =0   \label{eq:a5}.
\end{equation}
The Eq.~\eqref{eq:a5} is a bessel differential equation, for which a specific solution is obtained. Following this, the solution is incorporated into Eq.~\eqref{eq:appsub1}, and with the subsequent application of the Bessel function relations
\begin{subequations}
\begin{align}
\frac{d}{dx}\left( x^{v}J_{v}\left( x\right) \right)  &= x^{v}J_{v-1}\left(
x\right)  \\
\frac{d}{dx}\left( x^{-v}J_{v}\left( x\right) \right) 
&=-x^{-v}J_{v+1}\left( x\right).
\end{align}
\end{subequations}
The wavefunction is successfully derived
\begin{equation}
\psi =N\left( 
\begin{array}{c}
e^{i\left(j-\frac{\tau }{2}\right) \theta}\frac{2at}{2E-\Delta}\sqrt{\kappa} \varrho J_{(j-\frac{\tau }{2})}\left( 
\sqrt{\kappa}r\right)  \\ 
ie^{i\left( j+\frac{\tau }{2}\right) \theta } \varrho J_{(j+\frac{\tau)}{2}%
}\left(\sqrt{\kappa}r\right) 
\end{array}%
\right),
\end{equation}
where 
\begin{equation}
\varrho =\left\{ 
\begin{array}{c}
1 \\ 
\left( -1\right) ^{j+\frac{\tau }{2}}%
\end{array}%
\right. 
\begin{array}{c}
j+\frac{\tau }{2}\geq 0 \\ 
j+\frac{\tau }{2}<0%
\end{array}.
\end{equation}

\subsection{Nonzero magnetic field}

Continuing with the analysis of magnetic field flux within the QDs, we arrive at the following expressions 
\begin{subequations}
\begin{align}
at\left( \tau \frac{\partial }{\partial r}+\frac{1}{r}\left( j+\frac{\tau }{2%
}\right) +e\frac{B}{2\hbar }r\right) \chi _{2}\left( r\right)  &=\left( E-%
\frac{\Delta }{2}\right) \chi _{1}\left( r\right)  \\
-at\left( \tau \frac{\partial }{\partial r}-\frac{1}{r}\left( j-\frac{\tau }{%
2}\right) -e\frac{B}{2\hbar }r\right) \chi _{1}\left( r\right) &=\left(
E-\left( -\frac{\Delta }{2}+\lambda \tau s\right) \right) \chi _{2}\left(
r\right).
\end{align}
\end{subequations}
Adopting dimensionless units, we obtain
\begin{subequations}
\begin{align}
\left( \tau \frac{\partial }{\partial \rho }+\frac{1}{\rho }\left( j+\frac{%
\tau }{2}\right) +\beta \rho \right) \chi _{2}\left( \rho \right)  &=\left(
\varepsilon -\frac{1}{2}\delta \right) \chi _{1}\left( \rho \right) \label{eq:a9a} \\
\left( \tau \frac{\partial }{\partial \rho }-\frac{1}{\rho }\left( j-\frac{%
\tau }{2}\right) -\beta \rho \right) \chi _{1}\left( \rho \right) 
&=-\left( \varepsilon +\frac{1}{2}\delta -\Lambda \tau s\right) \chi
_{2}\left( \rho \right),\label{eq:a9b}
\end{align}
\end{subequations}
and we get $  -\frac{\left( \tau \frac{\partial }{\partial \rho }-\frac{1}{\rho }\left( j-
\frac{\tau }{2}\right) -\beta \rho \right) \chi _{1}\left( \rho \right) }{	\varepsilon +\frac{1}{2}\delta -\Lambda \tau s}=\chi _{2}\left( \rho \right)  $ 
from Eq.~\eqref{eq:a9b}, subsequently, that is substituted into Eq.~\eqref{eq:a9a}, yielding 
\begin{equation}
\frac{\partial }{\partial \rho ^{2}}\chi _{1}\left( \rho \right) -\left( j-%
\frac{\tau }{2}\right) ^{2}\frac{1}{\rho ^{2}}\chi _{1}\left( \rho \right) +%
\frac{1}{\rho }\frac{\partial }{\partial \rho }\chi _{1}\left( \rho \right)
-2\left( j+\frac{\tau }{2}\right) \beta \chi _{1}\left( \rho \right) -\beta
^{2}\rho ^{2}\chi _{1}\left( \rho \right) +\left( \varepsilon +\frac{1}{2}%
\delta -\Lambda \tau s\right) \left( \varepsilon -\frac{1}{2}\delta \right)
\chi _{1}\left( \rho \right) =0.
\end{equation}
We make the ansatz
$ 		   \chi _{1}\left( \rho \right) =\rho ^{\left\vert j-\frac{\tau }{2}\right\vert
}\exp \left( -\frac{\beta \rho ^{2}}{2}\right) \chi _{0}\left( \rho
^{2}\right) $ and define $x=\beta \rho ^{2}$, which enables us to derive
\begin{equation}
x\frac{\partial ^{2}}{\partial x^{2}}\chi _{0}\left( x\right) +\left( 
\left\vert j-\frac{\tau }{2}\right\vert +1 -x\right) \frac{\partial }{%
\partial x}\chi _{0}\left( x\right) -\left[ \frac{1}{2}\left( j+\frac{
\tau }{2} +\left\vert j-\frac{\tau }{2}\right\vert +1\right) -\frac{\left(\varepsilon +\frac{1}{2}\delta -\Lambda \tau s\right)
\left( \varepsilon -\frac{1}{2}\delta \right) }{4\beta }\right] \chi
_{0}\left( x\right) =0,
\end{equation}
where we introduce the definitions $ b =\left\vert j-\frac{\tau }{2}\right\vert +1 $ and 
$ \alpha  =\frac{1}{2}\left(  j+\frac{\tau }{2} +
\left\vert j-\frac{\tau }{2}\right\vert +1 \right) -\frac{\left(
\varepsilon +\frac{1}{2}\delta -\Lambda \tau s\right) \left( \varepsilon -%
\frac{1}{2}\delta \right) }{4\beta } $.
The resulting equation is identified as a confluent hypergeometric equation, from which the solution is deduced
$\chi _{1}\left( \rho \right) =C\rho ^{\left\vert j-\frac{\tau }{2}%
\right\vert }\exp \left( -\frac{\beta \rho ^{2}}{2}\right) \left.
_{1}F_{1}\right. \left( \alpha ,b,\beta \rho ^{2}\right).$ 
By incorporating another component into Eq.~\eqref{eq:a9b} and substituting it, we derive the wavefunction in the presence of a magnetic field
\begin{equation}
\psi =N\left( 
\begin{array}{c}
e^{i\left( j-\frac{\tau }{2}\right) \theta }\rho ^{\left\vert j-\frac{\tau }{%
2}\right\vert }\exp \left( -\frac{\beta \rho ^{2}}{2}\right) \left.
_{1}F_{1}\right. \left(\alpha,b,\beta \rho ^{2}\right)  \\ 
ie^{i\left( j+\frac{\tau }{2}\right) \theta }\rho ^{\left\vert j-\frac{\tau 
}{2}\right\vert -1}\exp \left( -\frac{\beta \rho ^{2}}{2}\right) \gamma \Gamma
\left( \rho \right) \label{eq:a12}
\end{array}%
\right),
\end{equation}
where we define $ \Gamma \left( \rho \right) = \xi \left. _{1}F_{1}\right. \left(\alpha,b,\beta \rho^{2}\right) +\tau \frac{2\alpha}{b}\beta \rho ^{2}\left. _{1}F_{1}\right. \left(\alpha+1,b+1,\beta \rho ^{2}\right)  $ with $ \xi =\tau \left\vert j-\frac{\tau }{2}\right\vert -j+\frac{\tau }{2}-\left(\tau +1\right) \beta \rho ^{2} $ and $ \gamma=-(\varepsilon +\frac{1}{2}\delta -\Lambda \tau s)^{-1} $.

\section{Coulomb interaction matrix elements}
\label{app:cou_mat}  

The Hamiltonian of the many-exciton system is given by
Eq. \eqref{many-bodyh}-\eqref{H_hh}. To numerically solve the
many-body Hamiltonian, it is essential to first derive the matrix elements
of the Coulomb interaction. 

The Coulomb interaction matrix elements is established by field operators 
\begin{eqnarray}
\Phi _{e}\left( \mathbf{r}_{e}\right)  &=&\sum_{s,n,j}\psi _{j,n,K,s}\left( 
\mathbf{r}_{e}\right) c_{j,n,s} \\
\Phi _{h}\left( \mathbf{r}_{h}\right)  &=&\sum_{n,j}\psi _{j,n,K,\downarrow
}\left( \mathbf{r}_{h}\right) d_{j,n},
\end{eqnarray}%
where we include the quantum numbers in the wave functions and operators,
including angular momentum $j$, principal quantum number $n$, electron spin
index $s$. We froze the spin of valence band to $\downarrow $ and suppose
the valley is polarized to $K$. By defining%
\begin{eqnarray*}
\Xi _{j_{i},n_{i},\tau _{i},s_{i}}^{\left( e\right) j_{k},n_{k},\tau
_{k},s_{k}}\left( \mathbf{q}\right)  &=&\int_{0}^{R_{e}}d\mathbf{r_{e}}\psi
_{j_{i},n_{i},\tau _{i},s_{i}}^{\ast }\left( \mathbf{r_{e}}\right) \psi
_{j_{k},n_{k},\tau _{k},s_{k}}\left( \mathbf{r_{e}}\right) e^{i\mathbf{%
q\cdot r_{e}}}, \\
\Xi _{j_{i},n_{i},\tau _{i},s_{i}}^{\left( h\right) j_{k},n_{k},\tau
_{k},s_{k}}\left( \mathbf{q}\right)  &=&\int_{0}^{R_{h}}d\mathbf{r_{h}}\psi
_{j_{i},n_{i},\tau _{i},s_{i}}^{\ast }\left( \mathbf{r_{h}}\right) \psi
_{j_{k},n_{k},\tau _{k},s_{k}}\left( \mathbf{r_{h}}\right) e^{i\mathbf{%
q\cdot r_{h}}},
\end{eqnarray*}
the interlayer electron-hole Coulomb interaction matrix element is
\begin{equation}
V_{j_{1},j_{2},j_{3},j_{4}}^{\left( eh\right) n_{1},n_{2},n_{3},n_{4}}= \int
qdqd\theta \frac{2\pi e^{2}}{\epsilon q}e^{-qd}\Xi
_{j_{1},n_{1},K,s}^{\left( e\right) j_{4},n_{4},K,s}\left( \mathbf{q}\right)
\Xi _{j_{2},n_{2},K,\downarrow }^{\left( h\right) j_{3},n_{3},K,\downarrow
}\left( -\mathbf{q}\right) ,
\label{eq:c3}
\end{equation}%
where $d$ is the interlayer separation. For intralayer Coulomb interaction $H_{ee}$, the electrons could have
different spin 
\begin{equation}
V_{j_{1},j_{2},j_{3},j_{4}}^{\left( ee\right) n_{1},n_{2},n_{3},n_{4}}=\int
qdqd\theta \frac{2\pi e^{2}}{\epsilon q}\Xi _{j_{1},n_{1},K,s}^{\left(
e\right) j_{4},n_{4},K,s}\left( \mathbf{q}\right) \Xi
_{j_{2},n_{2},K,s^{\prime }}^{\left( h\right) j_{3},n_{3},K,s^{\prime
}}\left( -\mathbf{q}\right) ,
\end{equation}%
and the hole-hole interaction term $H_{hh}$, the holes with spin up 
\begin{equation}
V_{j_{1},j_{2},j_{3},j_{4}}^{\left( hh\right) n_{1},n_{2},n_{3},n_{4}}=\int
qdqd\theta \frac{2\pi e^{2}}{\epsilon q}\Xi _{j_{1},n_{1},K,\downarrow
}^{\left( h\right) j_{4},n_{4},K,\downarrow }\left( \mathbf{q}\right) \Xi
_{j_{2},n_{2},K,\downarrow }^{\left( h\right) j_{3},n_{3},K,\downarrow
}\left( -\mathbf{q}\right) .  \nonumber
\end{equation}

\subsection{Zero magnetic field}

In the absence of a magnetic field, we initiate from Eq.~\eqref{eq:c3} to
derive the Coulomb interaction matrix element. And the derivation is
concerned with interlayer Coulomb interaction for the intralayer Coulomb
interaction can be numerically calculated by setting layer diatance $d=0$. Then we have 
\begin{eqnarray}
\Xi _{j_{i},n_{i},\tau _{i},s_{i}}^{\left( e\right) j_{k},n_{k},\tau
_{k},s_{k}}\left( \mathbf{q}\right)  &=&M_{j_{i},n_{i},\tau
_{i},s_{i}}^{j_{k},n_{k},\tau
_{k},s_{k}}\int_{0}^{R_{e}}r_{e}dr_{e}\int_{0}^{2\pi }d\theta _{e}e^{i%
\mathbf{q\cdot r_{e}}}  \nonumber \\
&&\times \left\{ C_{j_{i},n_{i},\tau _{i},s_{i}}^{j_{k},n_{k},\tau
_{k},s_{k}}e^{-i\left[ \left( j_{i}-\frac{\tau _{i}}{2}\right) -\left( j_{k}-%
\frac{\tau _{k}}{2}\right) \right] \theta _{e}}J_{\left( j_{i}-\frac{\tau
_{i}}{2}\right) }\left( \sqrt{\kappa _{j_{i},n_{i},\tau _{i},s_{i}}}%
r_{e}\right) J_{\left( j_{k}-\frac{\tau _{k}}{2}\right) }\left( \sqrt{\kappa
_{j_{k},n_{k},\tau _{k},s_{k}}}r_{e}\right) \right.   \nonumber \\
&&\left. +e^{-i\left[ \left( j_{i}+\frac{\tau _{i}}{2}\right) -\left( j_{k}+%
\frac{\tau _{k}}{2}\right) \right] \theta _{e}}J_{\left( j_{i}+\frac{\tau
_{i}}{2}\right) }\left( \sqrt{\kappa _{j_{i},n_{i},\tau _{i},s_{i}}}%
r_{e}\right) J_{\left( j_{k}+\frac{\tau _{k}}{2}\right) }\left( \sqrt{\kappa
_{j_{k},n_{k},\tau _{k},s_{k}}}r_{e}\right) \right\} ,  \label{eq:b7}
\end{eqnarray}%
where we define 
\begin{eqnarray}
C_{j_{i},n_{i},\tau _{i},s_{i}}^{j_{k},n_{k},\tau _{k},s_{k}} &=&\left(
2a_{e}t_{e}\right) ^{2}\frac{\sqrt{\kappa _{j_{i},n_{i},\tau _{i},s_{i}}}}{%
2E_{s_{i},\tau _{i},n_{i},j_{i}}-\Delta _{e}}\frac{\sqrt{\kappa
_{j_{k},n_{k},\tau _{k},s_{k}}}}{2E_{j_{k},n_{k},\tau _{k},s_{k}}-\Delta _{e}%
}, \\
M_{j_{i},n_{i},\tau _{i},s_{i}}^{j_{k},n_{k},\tau _{k},s_{k}}
&=&N_{j_{i},n_{i},\tau _{i},s_{i}}^{\ast }N_{j_{k},n_{k},\tau
_{k},s_{k}}\varrho _{j_{i},\tau _{i}}\varrho _{j_{k},\tau _{k}},
\end{eqnarray}%
with $a_e$ and $t_e$ being the lattice constant and the hopping parameter of the electron layer, respectively. Subsequently, we acquire
\begin{eqnarray}
\Xi _{j_{i},n_{i},\tau _{i},s_{i}}^{\left( e\right) j_{k},n_{k},\tau
_{k},s_{k}}\left( \mathbf{q}\right)  &=&M_{j_{i},n_{i},\tau
_{i},s_{i}}^{j_{k},n_{k},\tau _{k},s_{k}}R_{e}^{2}\int_{0}^{1}r_{e}^{\prime
}dr_{e}^{\prime }  \nonumber \\
&&\times \left\{ C_{j_{i},n_{i},\tau _{i},s_{i}}^{j_{k},n_{k},\tau
_{k},s_{k}}i^{\left\vert \left( j_{k}-\frac{\tau _{k}}{2}\right) -\left(
j_{i}-\frac{\tau _{i}}{2}\right) \right\vert }J_{\left\vert \left( j_{k}-%
\frac{\tau _{k}}{2}\right) -\left( j_{i}-\frac{\tau _{i}}{2}\right)
\right\vert }\left( q_{e}r_{e}^{\prime }\right) J_{\left( j_{i}-\frac{\tau
_{i}}{2}\right) }\left( K_{i}^{e}r_{e}^{\prime }\right) J_{\left( j_{k}-%
\frac{\tau _{k}}{2}\right) }\left( K_{k}^{e}r_{e}^{\prime }\right) \right.  
\nonumber \\
&&\left. +i^{\left\vert \left( j_{k}+\frac{\tau _{k}}{2}\right) -\left(
j_{i}+\frac{\tau _{i}}{2}\right) \right\vert }J_{\left\vert \left( j_{k}+%
\frac{\tau _{k}}{2}\right) -\left( j_{i}+\frac{\tau _{i}}{2}\right)
\right\vert }\left( q_{e}r_{e}^{\prime }\right) J_{\left( j_{i}+\frac{\tau
_{i}}{2}\right) }\left( K_{i}^{e}r_{e}^{\prime }\right) J_{\left( j_{k}+%
\frac{\tau _{k}}{2}\right) }\left( K_{k}^{e}r_{e}^{\prime }\right) \right\} .
\end{eqnarray}%
where we define$\quad K_{i}^{e}=R_{e}\sqrt{\kappa _{j_{i},n_{i},\tau
_{i},s_{i}}},\quad K_{k}^{e}=R_{e}\sqrt{\kappa _{j_{k},n_{k},\tau _{k},s_{k}}%
}$with dimensionless integral variables $q_{e}=R_{e}q,r_{e}^{\prime }=\frac{%
r_{e}}{R_{e}}.$ Similarly, we can obtain the function for hole 

\begin{eqnarray}
\Xi _{J_{i},N_{i},T_{i},S_{i}}^{\left( h\right)
J_{k},N_{k},T_{k},S_{k}}\left( \mathbf{q}\right) 
&=&M_{J_{i},N_{i},T_{i},S_{i}}^{J_{k},N_{k},T_{k},S_{k}}R_{h}^{2}%
\int_{0}^{1}r_{h}^{\prime }dr_{h}^{\prime }  \nonumber \\
&&\times \left\{
C_{J_{i},N_{i},T_{i},S_{i}}^{J_{k},N_{k},T_{k},S_{k}}i^{\left\vert \left(
J_{k}-\frac{T_{k}}{2}\right) -\left( J_{i}-\frac{T_{i}}{2}\right)
\right\vert }J_{\left\vert \left( J_{k}-\frac{T_{k}}{2}\right) -\left( J_{i}-%
\frac{T_{i}}{2}\right) \right\vert }\left( -q_{h}r_{h}^{\prime }\right)
J_{\left( J_{i}-\frac{T_{i}}{2}\right) }\left( K_{i}^{h}r_{h}^{\prime
}\right) J_{\left( j_{k}-\frac{\tau _{k}}{2}\right) }\left(
K_{k}^{h}r_{h}^{\prime }\right) \right.   \nonumber \\
&&\left. +i^{\left\vert \left( J_{k}+\frac{T_{k}}{2}\right) -\left( J_{i}+%
\frac{T_{i}}{2}\right) \right\vert }J_{\left\vert \left( J_{k}+\frac{T_{k}}{2%
}\right) -\left( J_{i}+\frac{T_{i}}{2}\right) \right\vert }\left(
-q_{h}r_{h}^{\prime }\right) J_{\left( J_{i}+\frac{T_{i}}{2}\right) }\left(
K_{i}^{h}r_{h}^{\prime }\right) J_{\left( J_{k}+\frac{T_{k}}{2}\right)
}\left( K_{k}^{h}r_{h}^{\prime }\right) \right\} 
\end{eqnarray}%
with$\quad K_{i}^{h}=R_{h}\sqrt{\kappa _{j_{i},n_{i},\tau _{i},s_{i}}}$ and $%
q_{h}=R_{h}q,r_{h}^{\prime }=\frac{r_{h}}{R_{h}}.$ We need to choose
appropriate dimensionless integral variables to improve the efficiency of
numerical integration. 

\subsection{With magnetic field}

In the presence of a magnetic field, only the function $\Xi $ needs to be
modified, 
\begin{eqnarray}
\Xi _{j_{i},n_{i},\tau _{i},s_{i}}^{\left( e\right) j_{k},n_{k},\tau
_{k},s_{k}}\left( \mathbf{q}\right)  &=&N_{j_{i},n_{i},\tau
_{i},s_{i}}^{\ast }N_{j_{k},n_{k},\tau _{k},s_{k}}\int_{0}^{\frac{R_{e}}{%
a_{e}}}\rho _{e}d\rho _{e}\int_{0}^{2\pi }d\theta _{e}e^{i\mathbf{q\cdot r}%
_{e}}\rho _{e}^{\left\vert j_{i}-\frac{\tau _{i}}{2}\right\vert +\left\vert
j_{k}-\frac{\tau _{k}}{2}\right\vert }e^{-\beta _{e}\rho _{e}^{2}}  \nonumber
\\
&&\times \left\{ e^{-i\left[ \left( j_{i}-\frac{\tau _{i}}{2}\right) -\left(
j_{k}-\frac{\tau _{k}}{2}\right) \right] \theta _{e}}\left. _{1}F_{1}\right.
\left( \alpha _{j_{i},n_{i},\tau _{i},s_{i}},b_{j_{i},\tau _{i}},\beta
_{e}\rho _{e}^{2}\right) \left. _{1}F_{1}\right. \left( \alpha
_{j_{k},n_{k},\tau _{k},s_{k}},b_{j_{k},\tau _{k}},\beta _{e}\rho
_{e}^{2}\right) \right.   \nonumber \\
&&\left. +e^{-i\left[ \left( j_{i}+\frac{\tau _{i}}{2}\right) -\left( j_{k}+%
\frac{\tau _{k}}{2}\right) \right] \theta _{e}}\gamma _{j_{i},n_{i},\tau
_{i},s_{i}}\gamma _{j_{k},n_{k},\tau _{k},s_{k}}\Gamma _{j_{i},n_{i},\tau
_{i},s_{i}}\left( \rho _{e}\right) \Gamma _{j_{k},n_{k},\tau
_{k},s_{k}}\left( \rho _{e}\right) \right\}   \nonumber \\
&=&N_{j_{i},n_{i},\tau _{i},s_{i}}^{\ast }N_{j_{k},n_{k},\tau
_{k},s_{k}}a_{L}^{e}i^{\left\vert \left( j_{k}-\frac{\tau _{k}}{2}\right)
-\left( j_{i}-\frac{\tau _{i}}{2}\right) \right\vert }\int dr_{e}^{\prime
}\left( a_{L}^{e}r_{e}^{\prime }\right) ^{\left\vert j_{i}-\frac{\tau _{i}}{2%
}\right\vert +\left\vert j_{k}-\frac{\tau _{k}}{2}\right\vert +1}\exp \left[
-\beta _{e}\left( a_{L}^{e}r_{e}^{\prime }\right) ^{2}\right]   \nonumber \\
&&\times J_{\left\vert \left( j_{k}-\frac{\tau _{k}}{2}\right) -\left( j_{i}-%
\frac{\tau _{i}}{2}\right) \right\vert }\left( q_{e}r_{e}^{\prime }\right)
\left. _{1}F_{1}\right. \left( \alpha _{j_{i},n_{i},\tau
_{i},s_{i}},b_{j_{i},\tau _{i}},\beta _{e}\left( a_{L}^{e}r_{e}^{\prime
}\right) ^{2}\right) \left. _{1}F_{1}\right. \left( \alpha
_{j_{k},n_{k},\tau _{k},s_{k}},b_{j_{k},\tau _{k}},\beta _{e}\left(
a_{L}^{e}r_{e}^{\prime }\right) ^{2}\right)   \nonumber \\
&&+N_{j_{i},n_{i},\tau _{i},s_{i}}^{\ast }N_{j_{k},n_{k},\tau
_{k},s_{k}}\gamma _{j_{i},n_{i},\tau _{i},s_{i}}\gamma _{j_{k},n_{k},\tau
_{k},s_{k}}i^{\left\vert \left( j_{k}+\frac{\tau _{k}}{2}\right) -\left(
j_{i}+\frac{\tau _{i}}{2}\right) \right\vert }a_{L}^{e}\int dr_{e}^{\prime
}\left( a_{L}^{e}r_{e}^{\prime }\right) ^{\left\vert j_{i}-\frac{\tau _{i}}{2%
}\right\vert +\left\vert j_{k}-\frac{\tau _{k}}{2}\right\vert -1}  \nonumber
\\
&&\times J_{\left\vert \left( j_{k}+\frac{\tau _{k}}{2}\right) -\left( j_{i}+%
\frac{\tau _{i}}{2}\right) \right\vert }\left( q_{e}r_{e}^{\prime }\right)
\exp \left[ -\beta _{e}\left( a_{L}^{e}r_{e}^{\prime }\right) ^{2}\right]
\Gamma _{j_{i},n_{i},\tau _{i},s_{i}}\left( a_{L}^{e}r_{e}^{\prime }\right)
\Gamma _{j_{k},n_{k},\tau _{k},s_{k}}\left( a_{L}^{e}r_{e}^{\prime }\right) 
\end{eqnarray}
where we define $\frac{R_{e}}{a_{e}}=a_{L}^{e}$, $r_{e}^{\prime }=\frac{r_{e}%
}{R_{e}}$ and $q_{e}=R_{e}q$. Here, the function $\Gamma \left( \rho \right) 
$ and $\gamma $ are defined in Eq.~\eqref{eq:a12}. Additionally, we expand
upon the established definitions and, similarly, we can deduce $\Xi
_{j_{i},n_{i},\tau _{i},s_{i}}^{\left( h\right) j_{k},n_{k},\tau _{k},s_{k}}$. 
Then all the Coulomb interaction matrix elements can be obtained.
\end{widetext}

\section*{Reference}
\bibliographystyle{apsrev4-2}
\bibliography{ref.bib}

\begin{thebibliography}{65}%
\makeatletter
\providecommand \@ifxundefined [1]{%
 \@ifx{#1\undefined}
}%
\providecommand \@ifnum [1]{%
 \ifnum #1\expandafter \@firstoftwo
 \else \expandafter \@secondoftwo
 \fi
}%
\providecommand \@ifx [1]{%
 \ifx #1\expandafter \@firstoftwo
 \else \expandafter \@secondoftwo
 \fi
}%
\providecommand \natexlab [1]{#1}%
\providecommand \enquote  [1]{``#1''}%
\providecommand \bibnamefont  [1]{#1}%
\providecommand \bibfnamefont [1]{#1}%
\providecommand \citenamefont [1]{#1}%
\providecommand \href@noop [0]{\@secondoftwo}%
\providecommand \href [0]{\begingroup \@sanitize@url \@href}%
\providecommand \@href[1]{\@@startlink{#1}\@@href}%
\providecommand \@@href[1]{\endgroup#1\@@endlink}%
\providecommand \@sanitize@url [0]{\catcode `\\12\catcode `\$12\catcode
  `\&12\catcode `\#12\catcode `\^12\catcode `\_12\catcode `\%12\relax}%
\providecommand \@@startlink[1]{}%
\providecommand \@@endlink[0]{}%
\providecommand \url  [0]{\begingroup\@sanitize@url \@url }%
\providecommand \@url [1]{\endgroup\@href {#1}{\urlprefix }}%
\providecommand \urlprefix  [0]{URL }%
\providecommand \Eprint [0]{\href }%
\providecommand \doibase [0]{https://doi.org/}%
\providecommand \selectlanguage [0]{\@gobble}%
\providecommand \bibinfo  [0]{\@secondoftwo}%
\providecommand \bibfield  [0]{\@secondoftwo}%
\providecommand \translation [1]{[#1]}%
\providecommand \BibitemOpen [0]{}%
\providecommand \bibitemStop [0]{}%
\providecommand \bibitemNoStop [0]{.\EOS\space}%
\providecommand \EOS [0]{\spacefactor3000\relax}%
\providecommand \BibitemShut  [1]{\csname bibitem#1\endcsname}%
\let\auto@bib@innerbib\@empty
\bibitem [{\citenamefont {Jiang}\ \emph {et~al.}(2021)\citenamefont {Jiang},
  \citenamefont {Chen}, \citenamefont {Zheng}, \citenamefont {Zheng},\ and\
  \citenamefont {Pan}}]{palinterlayerex}%
  \BibitemOpen
  \bibfield  {author} {\bibinfo {author} {\bibfnamefont {Y.}~\bibnamefont
  {Jiang}}, \bibinfo {author} {\bibfnamefont {S.}~\bibnamefont {Chen}},
  \bibinfo {author} {\bibfnamefont {W.}~\bibnamefont {Zheng}}, \bibinfo
  {author} {\bibfnamefont {B.}~\bibnamefont {Zheng}},\ and\ \bibinfo {author}
  {\bibfnamefont {A.}~\bibnamefont {Pan}},\ }\href
  {https://doi.org/https://doi.org/10.1038/s41377-021-00500-1} {\bibfield
  {journal} {\bibinfo  {journal} {Light Sci. Appl}\ }\textbf {\bibinfo {volume}
  {10}},\ \bibinfo {pages} {72} (\bibinfo {year} {2021})}\BibitemShut {NoStop}%
\bibitem [{\citenamefont {Latini}\ \emph {et~al.}(2017)\citenamefont {Latini},
  \citenamefont {Winther}, \citenamefont {Olsen},\ and\ \citenamefont
  {Thygesen}}]{latini2017interlayer}%
  \BibitemOpen
  \bibfield  {author} {\bibinfo {author} {\bibfnamefont {S.}~\bibnamefont
  {Latini}}, \bibinfo {author} {\bibfnamefont {K.~T.}\ \bibnamefont {Winther}},
  \bibinfo {author} {\bibfnamefont {T.}~\bibnamefont {Olsen}},\ and\ \bibinfo
  {author} {\bibfnamefont {K.~S.}\ \bibnamefont {Thygesen}},\ }\href
  {https://doi.org/https://doi.org/10.1021/acs.nanolett.6b04275} {\bibfield
  {journal} {\bibinfo  {journal} {Nano Lett.}\ }\textbf {\bibinfo {volume}
  {17}},\ \bibinfo {pages} {938} (\bibinfo {year} {2017})}\BibitemShut
  {NoStop}%
\bibitem [{\citenamefont {Manzeli}\ \emph {et~al.}(2017)\citenamefont
  {Manzeli}, \citenamefont {Ovchinnikov}, \citenamefont {Pasquier},
  \citenamefont {Yazyev},\ and\ \citenamefont {Kis}}]{2017app}%
  \BibitemOpen
  \bibfield  {author} {\bibinfo {author} {\bibfnamefont {S.}~\bibnamefont
  {Manzeli}}, \bibinfo {author} {\bibfnamefont {D.}~\bibnamefont
  {Ovchinnikov}}, \bibinfo {author} {\bibfnamefont {D.}~\bibnamefont
  {Pasquier}}, \bibinfo {author} {\bibfnamefont {O.~V.}\ \bibnamefont
  {Yazyev}},\ and\ \bibinfo {author} {\bibfnamefont {A.}~\bibnamefont {Kis}},\
  }\href {https://doi.org/https://doi.org/10.1038/natrevmats.2017.33}
  {\bibfield  {journal} {\bibinfo  {journal} {Nat. Rev. Mater}\ }\textbf
  {\bibinfo {volume} {2}},\ \bibinfo {pages} {1} (\bibinfo {year}
  {2017})}\BibitemShut {NoStop}%
\bibitem [{\citenamefont {Mueller}\ and\ \citenamefont
  {Malic}(2018)}]{2018app}%
  \BibitemOpen
  \bibfield  {author} {\bibinfo {author} {\bibfnamefont {T.}~\bibnamefont
  {Mueller}}\ and\ \bibinfo {author} {\bibfnamefont {E.}~\bibnamefont
  {Malic}},\ }\href {https://doi.org/https://doi.org/10.1038/s41699-018-0074-2}
  {\bibfield  {journal} {\bibinfo  {journal} {npj 2D Mater. Appl.}\ }\textbf
  {\bibinfo {volume} {2}},\ \bibinfo {pages} {29} (\bibinfo {year}
  {2018})}\BibitemShut {NoStop}%
\bibitem [{\citenamefont {Xiao}\ \emph {et~al.}(2012)\citenamefont {Xiao},
  \citenamefont {Liu}, \citenamefont {Feng}, \citenamefont {Xu},\ and\
  \citenamefont {Yao}}]{yao2012tmd}%
  \BibitemOpen
  \bibfield  {author} {\bibinfo {author} {\bibfnamefont {D.}~\bibnamefont
  {Xiao}}, \bibinfo {author} {\bibfnamefont {G.-B.}\ \bibnamefont {Liu}},
  \bibinfo {author} {\bibfnamefont {W.}~\bibnamefont {Feng}}, \bibinfo {author}
  {\bibfnamefont {X.}~\bibnamefont {Xu}},\ and\ \bibinfo {author}
  {\bibfnamefont {W.}~\bibnamefont {Yao}},\ }\href
  {https://doi.org/https://doi.org/10.1103/PhysRevLett.108.196802} {\bibfield
  {journal} {\bibinfo  {journal} {Phys. Rev. Lett.}\ }\textbf {\bibinfo
  {volume} {108}},\ \bibinfo {pages} {196802} (\bibinfo {year}
  {2012})}\BibitemShut {NoStop}%
\bibitem [{\citenamefont {Geim}\ and\ \citenamefont
  {Grigorieva}(2013)}]{geim2013van}%
  \BibitemOpen
  \bibfield  {author} {\bibinfo {author} {\bibfnamefont {A.~K.}\ \bibnamefont
  {Geim}}\ and\ \bibinfo {author} {\bibfnamefont {I.~V.}\ \bibnamefont
  {Grigorieva}},\ }\href {https://doi.org/https://doi.org/10.1038/nature12385}
  {\bibfield  {journal} {\bibinfo  {journal} {Nature}\ }\textbf {\bibinfo
  {volume} {499}},\ \bibinfo {pages} {419} (\bibinfo {year}
  {2013})}\BibitemShut {NoStop}%
\bibitem [{\citenamefont {Bellus}\ \emph {et~al.}(2017)\citenamefont {Bellus},
  \citenamefont {Li}, \citenamefont {Lane}, \citenamefont {Ceballos},
  \citenamefont {Cui}, \citenamefont {Zeng},\ and\ \citenamefont
  {Zhao}}]{bandtype}%
  \BibitemOpen
  \bibfield  {author} {\bibinfo {author} {\bibfnamefont {M.~Z.}\ \bibnamefont
  {Bellus}}, \bibinfo {author} {\bibfnamefont {M.}~\bibnamefont {Li}}, \bibinfo
  {author} {\bibfnamefont {S.~D.}\ \bibnamefont {Lane}}, \bibinfo {author}
  {\bibfnamefont {F.}~\bibnamefont {Ceballos}}, \bibinfo {author}
  {\bibfnamefont {Q.}~\bibnamefont {Cui}}, \bibinfo {author} {\bibfnamefont
  {X.~C.}\ \bibnamefont {Zeng}},\ and\ \bibinfo {author} {\bibfnamefont
  {H.}~\bibnamefont {Zhao}},\ }\href
  {https://doi.org/https://doi.org/10.1039/C6NH00144K} {\bibfield  {journal}
  {\bibinfo  {journal} {Nanoscale Horiz.}\ }\textbf {\bibinfo {volume} {2}},\
  \bibinfo {pages} {31} (\bibinfo {year} {2017})}\BibitemShut {NoStop}%
\bibitem [{\citenamefont {Kang}\ \emph {et~al.}(2013)\citenamefont {Kang},
  \citenamefont {Tongay}, \citenamefont {Zhou}, \citenamefont {Li},\ and\
  \citenamefont {Wu}}]{kang2013band}%
  \BibitemOpen
  \bibfield  {author} {\bibinfo {author} {\bibfnamefont {J.}~\bibnamefont
  {Kang}}, \bibinfo {author} {\bibfnamefont {S.}~\bibnamefont {Tongay}},
  \bibinfo {author} {\bibfnamefont {J.}~\bibnamefont {Zhou}}, \bibinfo {author}
  {\bibfnamefont {J.}~\bibnamefont {Li}},\ and\ \bibinfo {author}
  {\bibfnamefont {J.}~\bibnamefont {Wu}},\ }\href
  {https://doi.org/https://doi.org/10.1063/1.4774090} {\bibfield  {journal}
  {\bibinfo  {journal} {Appl. Phys. Lett}\ }\textbf {\bibinfo {volume} {102}},\
  \bibinfo {pages} {012111} (\bibinfo {year} {2013})}\BibitemShut {NoStop}%
\bibitem [{\citenamefont {Unuchek}\ \emph {et~al.}(2019)\citenamefont
  {Unuchek}, \citenamefont {Ciarrocchi}, \citenamefont {Avsar}, \citenamefont
  {Sun}, \citenamefont {Watanabe}, \citenamefont {Taniguchi},\ and\
  \citenamefont {Kis}}]{unuchek2019valley}%
  \BibitemOpen
  \bibfield  {author} {\bibinfo {author} {\bibfnamefont {D.}~\bibnamefont
  {Unuchek}}, \bibinfo {author} {\bibfnamefont {A.}~\bibnamefont {Ciarrocchi}},
  \bibinfo {author} {\bibfnamefont {A.}~\bibnamefont {Avsar}}, \bibinfo
  {author} {\bibfnamefont {Z.}~\bibnamefont {Sun}}, \bibinfo {author}
  {\bibfnamefont {K.}~\bibnamefont {Watanabe}}, \bibinfo {author}
  {\bibfnamefont {T.}~\bibnamefont {Taniguchi}},\ and\ \bibinfo {author}
  {\bibfnamefont {A.}~\bibnamefont {Kis}},\ }\href
  {https://doi.org/https://doi.org/10.1038/s41565-019-0559-y} {\bibfield
  {journal} {\bibinfo  {journal} {Nat.Nanotechnol.}\ }\textbf {\bibinfo
  {volume} {14}},\ \bibinfo {pages} {1104} (\bibinfo {year}
  {2019})}\BibitemShut {NoStop}%
\bibitem [{\citenamefont {Hajati}\ \emph {et~al.}(2023)\citenamefont {Hajati},
  \citenamefont {Alipourzadeh}, \citenamefont {Schulz},\ and\ \citenamefont
  {Berakdar}}]{hajati2023tuning}%
  \BibitemOpen
  \bibfield  {author} {\bibinfo {author} {\bibfnamefont {Y.}~\bibnamefont
  {Hajati}}, \bibinfo {author} {\bibfnamefont {M.}~\bibnamefont
  {Alipourzadeh}}, \bibinfo {author} {\bibfnamefont {D.}~\bibnamefont
  {Schulz}},\ and\ \bibinfo {author} {\bibfnamefont {J.}~\bibnamefont
  {Berakdar}},\ }\href
  {https://doi.org/https://doi.org/10.1103/PhysRevApplied.20.024075} {\bibfield
   {journal} {\bibinfo  {journal} {Phys. Rev. Appl.}\ }\textbf {\bibinfo
  {volume} {20}},\ \bibinfo {pages} {024075} (\bibinfo {year}
  {2023})}\BibitemShut {NoStop}%
\bibitem [{\citenamefont {Zhao}\ \emph {et~al.}(2022)\citenamefont {Zhao},
  \citenamefont {Du}, \citenamefont {Yang}, \citenamefont {Tian}, \citenamefont
  {Li}, \citenamefont {Shen}, \citenamefont {Tang}, \citenamefont {Chu},
  \citenamefont {Watanabe}, \citenamefont {Taniguchi} \emph
  {et~al.}}]{zhao2022interlayer}%
  \BibitemOpen
  \bibfield  {author} {\bibinfo {author} {\bibfnamefont {Y.}~\bibnamefont
  {Zhao}}, \bibinfo {author} {\bibfnamefont {L.}~\bibnamefont {Du}}, \bibinfo
  {author} {\bibfnamefont {S.}~\bibnamefont {Yang}}, \bibinfo {author}
  {\bibfnamefont {J.}~\bibnamefont {Tian}}, \bibinfo {author} {\bibfnamefont
  {X.}~\bibnamefont {Li}}, \bibinfo {author} {\bibfnamefont {C.}~\bibnamefont
  {Shen}}, \bibinfo {author} {\bibfnamefont {J.}~\bibnamefont {Tang}}, \bibinfo
  {author} {\bibfnamefont {Y.}~\bibnamefont {Chu}}, \bibinfo {author}
  {\bibfnamefont {K.}~\bibnamefont {Watanabe}}, \bibinfo {author}
  {\bibfnamefont {T.}~\bibnamefont {Taniguchi}}, \emph {et~al.},\ }\href
  {https://doi.org/https://doi.org/10.1103/PhysRevB.105.L041411} {\bibfield
  {journal} {\bibinfo  {journal} {Phys. Rev. B}\ }\textbf {\bibinfo {volume}
  {105}},\ \bibinfo {pages} {L041411} (\bibinfo {year} {2022})}\BibitemShut
  {NoStop}%
\bibitem [{\citenamefont {Rodin}\ and\ \citenamefont
  {Castro~Neto}(2013)}]{rodin2013excitonic}%
  \BibitemOpen
  \bibfield  {author} {\bibinfo {author} {\bibfnamefont {A.}~\bibnamefont
  {Rodin}}\ and\ \bibinfo {author} {\bibfnamefont {A.}~\bibnamefont
  {Castro~Neto}},\ }\href
  {https://doi.org/https://doi.org/10.1103/PhysRevB.88.195437} {\bibfield
  {journal} {\bibinfo  {journal} {Phys. Rev. B}\ }\textbf {\bibinfo {volume}
  {88}},\ \bibinfo {pages} {195437} (\bibinfo {year} {2013})}\BibitemShut
  {NoStop}%
\bibitem [{\citenamefont {Gartstein}\ \emph {et~al.}(2015)\citenamefont
  {Gartstein}, \citenamefont {Li},\ and\ \citenamefont
  {Zhang}}]{gartstein2015exciton}%
  \BibitemOpen
  \bibfield  {author} {\bibinfo {author} {\bibfnamefont {Y.~N.}\ \bibnamefont
  {Gartstein}}, \bibinfo {author} {\bibfnamefont {X.}~\bibnamefont {Li}},\ and\
  \bibinfo {author} {\bibfnamefont {C.}~\bibnamefont {Zhang}},\ }\href
  {https://doi.org/https://doi.org/10.1103/PhysRevB.92.075445} {\bibfield
  {journal} {\bibinfo  {journal} {Phys. Rev. B}\ }\textbf {\bibinfo {volume}
  {92}},\ \bibinfo {pages} {075445} (\bibinfo {year} {2015})}\BibitemShut
  {NoStop}%
\bibitem [{\citenamefont {Van~der Donck}\ and\ \citenamefont
  {Peeters}(2018)}]{van2018interlayer}%
  \BibitemOpen
  \bibfield  {author} {\bibinfo {author} {\bibfnamefont {M.}~\bibnamefont
  {Van~der Donck}}\ and\ \bibinfo {author} {\bibfnamefont {F.}~\bibnamefont
  {Peeters}},\ }\href
  {https://doi.org/https://doi.org/10.1103/PhysRevB.98.115104} {\bibfield
  {journal} {\bibinfo  {journal} {Phys. Rev. B}\ }\textbf {\bibinfo {volume}
  {98}},\ \bibinfo {pages} {115104} (\bibinfo {year} {2018})}\BibitemShut
  {NoStop}%
\bibitem [{\citenamefont {Wu}\ \emph {et~al.}(2019)\citenamefont {Wu},
  \citenamefont {Cheng},\ and\ \citenamefont {Wang}}]{wu2019exciton}%
  \BibitemOpen
  \bibfield  {author} {\bibinfo {author} {\bibfnamefont {S.}~\bibnamefont
  {Wu}}, \bibinfo {author} {\bibfnamefont {L.}~\bibnamefont {Cheng}},\ and\
  \bibinfo {author} {\bibfnamefont {Q.}~\bibnamefont {Wang}},\ }\href
  {https://doi.org/https://doi.org/10.1103/PhysRevB.100.115430} {\bibfield
  {journal} {\bibinfo  {journal} {Phys. Rev. B}\ }\textbf {\bibinfo {volume}
  {100}},\ \bibinfo {pages} {115430} (\bibinfo {year} {2019})}\BibitemShut
  {NoStop}%
\bibitem [{\citenamefont {Ten{\'o}rio}\ \emph {et~al.}(2023)\citenamefont
  {Ten{\'o}rio}, \citenamefont {Pereira}, \citenamefont {Mohseni},
  \citenamefont {Frederico}, \citenamefont {Hadizadeh}, \citenamefont
  {da~Costa},\ and\ \citenamefont {Chaves}}]{tuninterlayer2023}%
  \BibitemOpen
  \bibfield  {author} {\bibinfo {author} {\bibfnamefont {L.~G.}\ \bibnamefont
  {Ten{\'o}rio}}, \bibinfo {author} {\bibfnamefont {T.~A.}\ \bibnamefont
  {Pereira}}, \bibinfo {author} {\bibfnamefont {K.}~\bibnamefont {Mohseni}},
  \bibinfo {author} {\bibfnamefont {T.}~\bibnamefont {Frederico}}, \bibinfo
  {author} {\bibfnamefont {M.}~\bibnamefont {Hadizadeh}}, \bibinfo {author}
  {\bibfnamefont {D.~R.}\ \bibnamefont {da~Costa}},\ and\ \bibinfo {author}
  {\bibfnamefont {A.~J.}\ \bibnamefont {Chaves}},\ }\href
  {https://doi.org/https://doi.org/10.1103/PhysRevB.108.035421} {\bibfield
  {journal} {\bibinfo  {journal} {Phys. Rev. B}\ }\textbf {\bibinfo {volume}
  {108}},\ \bibinfo {pages} {035421} (\bibinfo {year} {2023})}\BibitemShut
  {NoStop}%
\bibitem [{\citenamefont {Su}\ and\ \citenamefont
  {MacDonald}(2008)}]{su2008make}%
  \BibitemOpen
  \bibfield  {author} {\bibinfo {author} {\bibfnamefont {J.-J.}\ \bibnamefont
  {Su}}\ and\ \bibinfo {author} {\bibfnamefont {A.}~\bibnamefont {MacDonald}},\
  }\href {https://doi.org/https://doi.org/10.1038/nphys1055} {\bibfield
  {journal} {\bibinfo  {journal} {Nat. Phys.}\ }\textbf {\bibinfo {volume}
  {4}},\ \bibinfo {pages} {799} (\bibinfo {year} {2008})}\BibitemShut {NoStop}%
\bibitem [{\citenamefont {Dai}\ and\ \citenamefont {Fu}(2024)}]{fu2024bilayer}%
  \BibitemOpen
  \bibfield  {author} {\bibinfo {author} {\bibfnamefont {D.~D.}\ \bibnamefont
  {Dai}}\ and\ \bibinfo {author} {\bibfnamefont {L.}~\bibnamefont {Fu}},\
  }\href {https://doi.org/https://doi.org/10.1103/PhysRevLett.132.196202}
  {\bibfield  {journal} {\bibinfo  {journal} {Phys. Rev. Lett.}\ }\textbf
  {\bibinfo {volume} {132}},\ \bibinfo {pages} {196202} (\bibinfo {year}
  {2024})}\BibitemShut {NoStop}%
\bibitem [{\citenamefont {Berman}\ and\ \citenamefont
  {Kezerashvili}(2017)}]{2017superfluidity}%
  \BibitemOpen
  \bibfield  {author} {\bibinfo {author} {\bibfnamefont {O.~L.}\ \bibnamefont
  {Berman}}\ and\ \bibinfo {author} {\bibfnamefont {R.~Y.}\ \bibnamefont
  {Kezerashvili}},\ }\href
  {https://doi.org/https://doi.org/10.1103/PhysRevB.96.094502} {\bibfield
  {journal} {\bibinfo  {journal} {Phys. Rev. B}\ }\textbf {\bibinfo {volume}
  {96}},\ \bibinfo {pages} {094502} (\bibinfo {year} {2017})}\BibitemShut
  {NoStop}%
\bibitem [{\citenamefont {Abouelkomsan}\ \emph {et~al.}(2024)\citenamefont
  {Abouelkomsan}, \citenamefont {Bergholtz},\ and\ \citenamefont
  {Chatterjee}}]{2024multiferroicityTMDS}%
  \BibitemOpen
  \bibfield  {author} {\bibinfo {author} {\bibfnamefont {A.}~\bibnamefont
  {Abouelkomsan}}, \bibinfo {author} {\bibfnamefont {E.~J.}\ \bibnamefont
  {Bergholtz}},\ and\ \bibinfo {author} {\bibfnamefont {S.}~\bibnamefont
  {Chatterjee}},\ }\href
  {https://doi.org/https://doi.org/10.1103/PhysRevLett.133.026801} {\bibfield
  {journal} {\bibinfo  {journal} {Phys. Rev. Lett.}\ }\textbf {\bibinfo
  {volume} {133}},\ \bibinfo {pages} {026801} (\bibinfo {year}
  {2024})}\BibitemShut {NoStop}%
\bibitem [{\citenamefont {Yang}\ \emph {et~al.}(2023)\citenamefont {Yang},
  \citenamefont {Zhang}, \citenamefont {Zhang}, \citenamefont {Lin},
  \citenamefont {Liu}, \citenamefont {Huang}, \citenamefont {Zhang},
  \citenamefont {Luo}, \citenamefont {He},\ and\ \citenamefont
  {Yuan}}]{yuan2023controlled}%
  \BibitemOpen
  \bibfield  {author} {\bibinfo {author} {\bibfnamefont {X.}~\bibnamefont
  {Yang}}, \bibinfo {author} {\bibfnamefont {S.}~\bibnamefont {Zhang}},
  \bibinfo {author} {\bibfnamefont {Z.}~\bibnamefont {Zhang}}, \bibinfo
  {author} {\bibfnamefont {J.}~\bibnamefont {Lin}}, \bibinfo {author}
  {\bibfnamefont {X.}~\bibnamefont {Liu}}, \bibinfo {author} {\bibfnamefont
  {Z.}~\bibnamefont {Huang}}, \bibinfo {author} {\bibfnamefont
  {L.}~\bibnamefont {Zhang}}, \bibinfo {author} {\bibfnamefont
  {W.}~\bibnamefont {Luo}}, \bibinfo {author} {\bibfnamefont {J.}~\bibnamefont
  {He}},\ and\ \bibinfo {author} {\bibfnamefont {X.}~\bibnamefont {Yuan}},\
  }\href {https://doi.org/https://doi.org/10.1016/j.physe.2023.115788}
  {\bibfield  {journal} {\bibinfo  {journal} {Physica E}\ }\textbf {\bibinfo
  {volume} {153}},\ \bibinfo {pages} {115788} (\bibinfo {year}
  {2023})}\BibitemShut {NoStop}%
\bibitem [{\citenamefont {Chen}\ \emph {et~al.}(2024)\citenamefont {Chen},
  \citenamefont {Chang}, \citenamefont {Yang}, \citenamefont {Lu},
  \citenamefont {Zhang}, \citenamefont {Zhang}, \citenamefont {He},\ and\
  \citenamefont {Yuan}}]{yuan2024vdw}%
  \BibitemOpen
  \bibfield  {author} {\bibinfo {author} {\bibfnamefont {M.}~\bibnamefont
  {Chen}}, \bibinfo {author} {\bibfnamefont {R.}~\bibnamefont {Chang}},
  \bibinfo {author} {\bibfnamefont {X.}~\bibnamefont {Yang}}, \bibinfo {author}
  {\bibfnamefont {C.}~\bibnamefont {Lu}}, \bibinfo {author} {\bibfnamefont
  {S.}~\bibnamefont {Zhang}}, \bibinfo {author} {\bibfnamefont
  {Z.}~\bibnamefont {Zhang}}, \bibinfo {author} {\bibfnamefont
  {J.}~\bibnamefont {He}},\ and\ \bibinfo {author} {\bibfnamefont
  {X.}~\bibnamefont {Yuan}},\ }\href {https://doi.org/10.1088/1361-6463/ad30ae}
  {\bibfield  {journal} {\bibinfo  {journal} {J. Phys. D}\ }\textbf {\bibinfo
  {volume} {57}},\ \bibinfo {pages} {235103} (\bibinfo {year}
  {2024})}\BibitemShut {NoStop}%
\bibitem [{\citenamefont {Lu}\ \emph {et~al.}(2024)\citenamefont {Lu},
  \citenamefont {Zhang}, \citenamefont {Chen}, \citenamefont {Chen},
  \citenamefont {Zhu}, \citenamefont {Zhang}, \citenamefont {He}, \citenamefont
  {Zhang},\ and\ \citenamefont {Yuan}}]{lu2024van}%
  \BibitemOpen
  \bibfield  {author} {\bibinfo {author} {\bibfnamefont {C.}~\bibnamefont
  {Lu}}, \bibinfo {author} {\bibfnamefont {S.}~\bibnamefont {Zhang}}, \bibinfo
  {author} {\bibfnamefont {M.}~\bibnamefont {Chen}}, \bibinfo {author}
  {\bibfnamefont {H.}~\bibnamefont {Chen}}, \bibinfo {author} {\bibfnamefont
  {M.}~\bibnamefont {Zhu}}, \bibinfo {author} {\bibfnamefont {Z.}~\bibnamefont
  {Zhang}}, \bibinfo {author} {\bibfnamefont {J.}~\bibnamefont {He}}, \bibinfo
  {author} {\bibfnamefont {L.}~\bibnamefont {Zhang}},\ and\ \bibinfo {author}
  {\bibfnamefont {X.}~\bibnamefont {Yuan}},\ }\href
  {https://doi.org/https://doi.org/10.1007/s11467-024-1404-9} {\bibfield
  {journal} {\bibinfo  {journal} {Front. Phys.}\ }\textbf {\bibinfo {volume}
  {19}},\ \bibinfo {pages} {53206} (\bibinfo {year} {2024})}\BibitemShut
  {NoStop}%
\bibitem [{\citenamefont {Chakraborty}(1999)}]{tapsh1999129}%
  \BibitemOpen
  \bibfield  {author} {\bibinfo {author} {\bibfnamefont {T.}~\bibnamefont
  {Chakraborty}},\ }\href
  {https://doi.org/https://doi.org/10.1016/B978-0-444-50258-2.X5000-7} {\emph
  {\bibinfo {title} {Quantum Dots}}}\ (\bibinfo  {publisher} {North-Holland},\
  \bibinfo {address} {Amsterdam},\ \bibinfo {year} {1999})\BibitemShut
  {NoStop}%
\bibitem [{\citenamefont {Chakraborty}(2003)}]{chakraborty2003nanoscopic}%
  \BibitemOpen
  \bibfield  {author} {\bibinfo {author} {\bibfnamefont {T.}~\bibnamefont
  {Chakraborty}},\ }\href
  {https://doi.org/https://doi.org/10.1007/978-3-540-44838-9_6} {\bibfield
  {journal} {\bibinfo  {journal} {Adv. Sol. State Phy.}\ }\textbf {\bibinfo
  {volume} {43}},\ \bibinfo {pages} {79} (\bibinfo {year} {2003})}\BibitemShut
  {NoStop}%
\bibitem [{\citenamefont {Chakraborty}\ \emph
  {et~al.}(2018{\natexlab{a}})\citenamefont {Chakraborty}, \citenamefont
  {Manaselyan},\ and\ \citenamefont {Barseghyan}}]{tapshringbook}%
  \BibitemOpen
  \bibfield  {author} {\bibinfo {author} {\bibfnamefont {T.}~\bibnamefont
  {Chakraborty}}, \bibinfo {author} {\bibfnamefont {A.~K.}\ \bibnamefont
  {Manaselyan}},\ and\ \bibinfo {author} {\bibfnamefont {M.~G.}\ \bibnamefont
  {Barseghyan}},\ }in\ \href
  {https://doi.org/https://doi.org/10.1007/978-3-319-95159-1_11} {\emph
  {\bibinfo {booktitle} {Physics of Quantum Rings}}},\ \bibinfo {editor}
  {edited by\ \bibinfo {editor} {\bibfnamefont {V.~M.}\ \bibnamefont {Fomin}}}\
  (\bibinfo  {publisher} {Springer},\ \bibinfo {year} {2018})\ pp.\ \bibinfo
  {pages} {283--326}\BibitemShut {NoStop}%
\bibitem [{\citenamefont {Chen}\ \emph {et~al.}(2007)\citenamefont {Chen},
  \citenamefont {Apalkov},\ and\ \citenamefont {Chakraborty}}]{chen2007fock}%
  \BibitemOpen
  \bibfield  {author} {\bibinfo {author} {\bibfnamefont {H.-Y.}\ \bibnamefont
  {Chen}}, \bibinfo {author} {\bibfnamefont {V.}~\bibnamefont {Apalkov}},\ and\
  \bibinfo {author} {\bibfnamefont {T.}~\bibnamefont {Chakraborty}},\ }\href
  {https://doi.org/https://doi.org/10.1103/PhysRevLett.98.186803} {\bibfield
  {journal} {\bibinfo  {journal} {Phys. Rev. Lett.}\ }\textbf {\bibinfo
  {volume} {98}},\ \bibinfo {pages} {186803} (\bibinfo {year}
  {2007})}\BibitemShut {NoStop}%
\bibitem [{\citenamefont {Halonen}\ \emph {et~al.}(1992)\citenamefont
  {Halonen}, \citenamefont {Chakraborty},\ and\ \citenamefont
  {Pietil{\"a}inen}}]{tapshexciton}%
  \BibitemOpen
  \bibfield  {author} {\bibinfo {author} {\bibfnamefont {V.}~\bibnamefont
  {Halonen}}, \bibinfo {author} {\bibfnamefont {T.}~\bibnamefont
  {Chakraborty}},\ and\ \bibinfo {author} {\bibfnamefont {P.}~\bibnamefont
  {Pietil{\"a}inen}},\ }\href
  {https://doi.org/https://doi.org/10.1103/PhysRevB.45.5980} {\bibfield
  {journal} {\bibinfo  {journal} {Phys. Rev. B}\ }\textbf {\bibinfo {volume}
  {45}},\ \bibinfo {pages} {5980} (\bibinfo {year} {1992})}\BibitemShut
  {NoStop}%
\bibitem [{\citenamefont {Zrenner}\ \emph {et~al.}(1994)\citenamefont
  {Zrenner}, \citenamefont {Butov}, \citenamefont {Hagn}, \citenamefont
  {Abstreiter}, \citenamefont {B{\"o}hm},\ and\ \citenamefont
  {Weimann}}]{zrenner1994quantum}%
  \BibitemOpen
  \bibfield  {author} {\bibinfo {author} {\bibfnamefont {A.}~\bibnamefont
  {Zrenner}}, \bibinfo {author} {\bibfnamefont {L.}~\bibnamefont {Butov}},
  \bibinfo {author} {\bibfnamefont {M.}~\bibnamefont {Hagn}}, \bibinfo {author}
  {\bibfnamefont {G.}~\bibnamefont {Abstreiter}}, \bibinfo {author}
  {\bibfnamefont {G.}~\bibnamefont {B{\"o}hm}},\ and\ \bibinfo {author}
  {\bibfnamefont {G.}~\bibnamefont {Weimann}},\ }\href
  {https://doi.org/https://doi.org/10.1103/PhysRevLett.72.3382} {\bibfield
  {journal} {\bibinfo  {journal} {Phys. Rev. Lett.}\ }\textbf {\bibinfo
  {volume} {72}},\ \bibinfo {pages} {3382} (\bibinfo {year}
  {1994})}\BibitemShut {NoStop}%
\bibitem [{\citenamefont {Zeng}\ and\ \citenamefont
  {MacDonald}(2022)}]{mac2022strong}%
  \BibitemOpen
  \bibfield  {author} {\bibinfo {author} {\bibfnamefont {Y.}~\bibnamefont
  {Zeng}}\ and\ \bibinfo {author} {\bibfnamefont {A.~H.}\ \bibnamefont
  {MacDonald}},\ }\href
  {https://doi.org/https://doi.org/10.1103/PhysRevB.106.035115} {\bibfield
  {journal} {\bibinfo  {journal} {Phys. Rev. B}\ }\textbf {\bibinfo {volume}
  {106}},\ \bibinfo {pages} {035115} (\bibinfo {year} {2022})}\BibitemShut
  {NoStop}%
\bibitem [{\citenamefont {Zarenia}\ \emph {et~al.}(2013)\citenamefont
  {Zarenia}, \citenamefont {Partoens}, \citenamefont {Chakraborty},\ and\
  \citenamefont {Peeters}}]{tapsh2013electron}%
  \BibitemOpen
  \bibfield  {author} {\bibinfo {author} {\bibfnamefont {M.}~\bibnamefont
  {Zarenia}}, \bibinfo {author} {\bibfnamefont {B.}~\bibnamefont {Partoens}},
  \bibinfo {author} {\bibfnamefont {T.}~\bibnamefont {Chakraborty}},\ and\
  \bibinfo {author} {\bibfnamefont {F.}~\bibnamefont {Peeters}},\ }\href
  {https://doi.org/https://doi.org/10.1103/PhysRevB.88.245432} {\bibfield
  {journal} {\bibinfo  {journal} {Phys. Rev. B}\ }\textbf {\bibinfo {volume}
  {88}},\ \bibinfo {pages} {245432} (\bibinfo {year} {2013})}\BibitemShut
  {NoStop}%
\bibitem [{\citenamefont {Jing}\ \emph {et~al.}(2022)\citenamefont {Jing},
  \citenamefont {Zhang}, \citenamefont {Qin}, \citenamefont {Luo},
  \citenamefont {Cao}, \citenamefont {Li}, \citenamefont {Song},\ and\
  \citenamefont {Guo}}]{2022gateQD}%
  \BibitemOpen
  \bibfield  {author} {\bibinfo {author} {\bibfnamefont {F.-M.}\ \bibnamefont
  {Jing}}, \bibinfo {author} {\bibfnamefont {Z.-Z.}\ \bibnamefont {Zhang}},
  \bibinfo {author} {\bibfnamefont {G.-Q.}\ \bibnamefont {Qin}}, \bibinfo
  {author} {\bibfnamefont {G.}~\bibnamefont {Luo}}, \bibinfo {author}
  {\bibfnamefont {G.}~\bibnamefont {Cao}}, \bibinfo {author} {\bibfnamefont
  {H.-O.}\ \bibnamefont {Li}}, \bibinfo {author} {\bibfnamefont {X.-X.}\
  \bibnamefont {Song}},\ and\ \bibinfo {author} {\bibfnamefont {G.-P.}\
  \bibnamefont {Guo}},\ }\href
  {https://doi.org/https://doi.org/10.1002/qute.202100162} {\bibfield
  {journal} {\bibinfo  {journal} {Adv. Quantum Technol.}\ }\textbf {\bibinfo
  {volume} {5}},\ \bibinfo {pages} {2100162} (\bibinfo {year}
  {2022})}\BibitemShut {NoStop}%
\bibitem [{\citenamefont {Wang}\ \emph {et~al.}(2019)\citenamefont {Wang},
  \citenamefont {Yuan}, \citenamefont {Liu}, \citenamefont {Sun}, \citenamefont
  {Muruganathan},\ and\ \citenamefont {Mizuta}}]{js1}%
  \BibitemOpen
  \bibfield  {author} {\bibinfo {author} {\bibfnamefont {Z.}~\bibnamefont
  {Wang}}, \bibinfo {author} {\bibfnamefont {Y.}~\bibnamefont {Yuan}}, \bibinfo
  {author} {\bibfnamefont {X.}~\bibnamefont {Liu}}, \bibinfo {author}
  {\bibfnamefont {J.}~\bibnamefont {Sun}}, \bibinfo {author} {\bibfnamefont
  {M.}~\bibnamefont {Muruganathan}},\ and\ \bibinfo {author} {\bibfnamefont
  {H.}~\bibnamefont {Mizuta}},\ }\href
  {https://doi.org/10.1021/acsnano.9b02935} {\bibfield  {journal} {\bibinfo
  {journal} {ACS Nano}\ }\textbf {\bibinfo {volume} {13}},\ \bibinfo {pages}
  {7502} (\bibinfo {year} {2019})}\BibitemShut {NoStop}%
\bibitem [{\citenamefont {Wang}\ \emph {et~al.}(2021)\citenamefont {Wang},
  \citenamefont {Yuan}, \citenamefont {Liu}, \citenamefont {Muruganathan},
  \citenamefont {Mizuta},\ and\ \citenamefont {Sun}}]{js2}%
  \BibitemOpen
  \bibfield  {author} {\bibinfo {author} {\bibfnamefont {Z.}~\bibnamefont
  {Wang}}, \bibinfo {author} {\bibfnamefont {Y.}~\bibnamefont {Yuan}}, \bibinfo
  {author} {\bibfnamefont {X.}~\bibnamefont {Liu}}, \bibinfo {author}
  {\bibfnamefont {M.}~\bibnamefont {Muruganathan}}, \bibinfo {author}
  {\bibfnamefont {H.}~\bibnamefont {Mizuta}},\ and\ \bibinfo {author}
  {\bibfnamefont {J.}~\bibnamefont {Sun}},\ }\href
  {https://doi.org/10.1063/5.0038419} {\bibfield  {journal} {\bibinfo
  {journal} {Appl. Phys. Lett}\ }\textbf {\bibinfo {volume} {118}},\ \bibinfo
  {pages} {083105} (\bibinfo {year} {2021})}\BibitemShut {NoStop}%
\bibitem [{\citenamefont {Guo}\ \emph {et~al.}(2019)\citenamefont {Guo},
  \citenamefont {Shu}, \citenamefont {Dong},\ and\ \citenamefont {Nori}}]{shu}%
  \BibitemOpen
  \bibfield  {author} {\bibinfo {author} {\bibfnamefont {Y.}~\bibnamefont
  {Guo}}, \bibinfo {author} {\bibfnamefont {C.-C.}\ \bibnamefont {Shu}},
  \bibinfo {author} {\bibfnamefont {D.}~\bibnamefont {Dong}},\ and\ \bibinfo
  {author} {\bibfnamefont {F.}~\bibnamefont {Nori}},\ }\href
  {https://doi.org/10.1103/PhysRevLett.123.223202} {\bibfield  {journal}
  {\bibinfo  {journal} {Phys. Rev. Lett.}\ }\textbf {\bibinfo {volume} {123}},\
  \bibinfo {pages} {223202} (\bibinfo {year} {2019})}\BibitemShut {NoStop}%
\bibitem [{\citenamefont {Shou}\ \emph {et~al.}(2021)\citenamefont {Shou},
  \citenamefont {Zhang}, \citenamefont {Luo},\ and\ \citenamefont
  {Huang}}]{luooe}%
  \BibitemOpen
  \bibfield  {author} {\bibinfo {author} {\bibfnamefont {C.}~\bibnamefont
  {Shou}}, \bibinfo {author} {\bibfnamefont {Q.}~\bibnamefont {Zhang}},
  \bibinfo {author} {\bibfnamefont {W.}~\bibnamefont {Luo}},\ and\ \bibinfo
  {author} {\bibfnamefont {G.}~\bibnamefont {Huang}},\ }\href
  {https://doi.org/10.1364/OE.416791} {\bibfield  {journal} {\bibinfo
  {journal} {Opt. Express}\ }\textbf {\bibinfo {volume} {29}},\ \bibinfo
  {pages} {9772} (\bibinfo {year} {2021})}\BibitemShut {NoStop}%
\bibitem [{\citenamefont {Mirzakhani}\ \emph {et~al.}(2016)\citenamefont
  {Mirzakhani}, \citenamefont {Zarenia}, \citenamefont {Da~Costa},
  \citenamefont {Ketabi},\ and\ \citenamefont
  {Peeters}}]{peetertrilayercomsol}%
  \BibitemOpen
  \bibfield  {author} {\bibinfo {author} {\bibfnamefont {M.}~\bibnamefont
  {Mirzakhani}}, \bibinfo {author} {\bibfnamefont {M.}~\bibnamefont {Zarenia}},
  \bibinfo {author} {\bibfnamefont {D.}~\bibnamefont {Da~Costa}}, \bibinfo
  {author} {\bibfnamefont {S.}~\bibnamefont {Ketabi}},\ and\ \bibinfo {author}
  {\bibfnamefont {F.}~\bibnamefont {Peeters}},\ }\href
  {https://doi.org/https://doi.org/10.1103/PhysRevB.94.165423} {\bibfield
  {journal} {\bibinfo  {journal} {Phys. Rev. B}\ }\textbf {\bibinfo {volume}
  {94}},\ \bibinfo {pages} {165423} (\bibinfo {year} {2016})}\BibitemShut
  {NoStop}%
\bibitem [{\citenamefont {Korm{\'a}nyos}\ \emph {et~al.}(2014)\citenamefont
  {Korm{\'a}nyos}, \citenamefont {Z{\'o}lyomi}, \citenamefont {Drummond},\ and\
  \citenamefont {Burkard}}]{kormanyos2014spin}%
  \BibitemOpen
  \bibfield  {author} {\bibinfo {author} {\bibfnamefont {A.}~\bibnamefont
  {Korm{\'a}nyos}}, \bibinfo {author} {\bibfnamefont {V.}~\bibnamefont
  {Z{\'o}lyomi}}, \bibinfo {author} {\bibfnamefont {N.~D.}\ \bibnamefont
  {Drummond}},\ and\ \bibinfo {author} {\bibfnamefont {G.}~\bibnamefont
  {Burkard}},\ }\href
  {https://doi.org/https://doi.org/10.1103/PhysRevX.4.011034} {\bibfield
  {journal} {\bibinfo  {journal} {Phys. Rev. X}\ }\textbf {\bibinfo {volume}
  {4}},\ \bibinfo {pages} {011034} (\bibinfo {year} {2014})}\BibitemShut
  {NoStop}%
\bibitem [{\citenamefont {Liu}\ \emph {et~al.}(2014)\citenamefont {Liu},
  \citenamefont {Pang}, \citenamefont {Yao},\ and\ \citenamefont
  {Yao}}]{liu2014intervalley}%
  \BibitemOpen
  \bibfield  {author} {\bibinfo {author} {\bibfnamefont {G.-B.}\ \bibnamefont
  {Liu}}, \bibinfo {author} {\bibfnamefont {H.}~\bibnamefont {Pang}}, \bibinfo
  {author} {\bibfnamefont {Y.}~\bibnamefont {Yao}},\ and\ \bibinfo {author}
  {\bibfnamefont {W.}~\bibnamefont {Yao}},\ }\href
  {https://doi.org/10.1088/1367-2630/16/10/105011} {\bibfield  {journal}
  {\bibinfo  {journal} {New J. Phys.}\ }\textbf {\bibinfo {volume} {16}},\
  \bibinfo {pages} {105011} (\bibinfo {year} {2014})}\BibitemShut {NoStop}%
\bibitem [{\citenamefont {Qu}\ \emph {et~al.}(2017)\citenamefont {Qu},
  \citenamefont {Dias}, \citenamefont {Fu}, \citenamefont {Villegas-Lelovsky},\
  and\ \citenamefont {Azevedo}}]{qu2017tunable}%
  \BibitemOpen
  \bibfield  {author} {\bibinfo {author} {\bibfnamefont {F.}~\bibnamefont
  {Qu}}, \bibinfo {author} {\bibfnamefont {A.}~\bibnamefont {Dias}}, \bibinfo
  {author} {\bibfnamefont {J.}~\bibnamefont {Fu}}, \bibinfo {author}
  {\bibfnamefont {L.}~\bibnamefont {Villegas-Lelovsky}},\ and\ \bibinfo
  {author} {\bibfnamefont {D.~L.}\ \bibnamefont {Azevedo}},\ }\href
  {https://doi.org/https://doi.org/10.1038/srep41044} {\bibfield  {journal}
  {\bibinfo  {journal} {Sci. Rep.}\ }\textbf {\bibinfo {volume} {7}},\ \bibinfo
  {pages} {41044} (\bibinfo {year} {2017})}\BibitemShut {NoStop}%
\bibitem [{\citenamefont {Dias}\ \emph {et~al.}(2016)\citenamefont {Dias},
  \citenamefont {Fu}, \citenamefont {Villegas-Lelovsky},\ and\ \citenamefont
  {Qu}}]{qu2016robust}%
  \BibitemOpen
  \bibfield  {author} {\bibinfo {author} {\bibfnamefont {A.}~\bibnamefont
  {Dias}}, \bibinfo {author} {\bibfnamefont {J.}~\bibnamefont {Fu}}, \bibinfo
  {author} {\bibfnamefont {L.}~\bibnamefont {Villegas-Lelovsky}},\ and\
  \bibinfo {author} {\bibfnamefont {F.}~\bibnamefont {Qu}},\ }\href
  {https://doi.org/10.1088/0953-8984/28/37/375803} {\bibfield  {journal}
  {\bibinfo  {journal} {J. Condens. Matter Phys.}\ }\textbf {\bibinfo {volume}
  {28}},\ \bibinfo {pages} {375803} (\bibinfo {year} {2016})}\BibitemShut
  {NoStop}%
\bibitem [{\citenamefont {Miravet}\ \emph {et~al.}(2023)\citenamefont
  {Miravet}, \citenamefont {Alt{\i}nta{\c{s}}}, \citenamefont {Rodrigues},
  \citenamefont {Bieniek}, \citenamefont {Korkusinski},\ and\ \citenamefont
  {Hawrylak}}]{miravet2023interacting}%
  \BibitemOpen
  \bibfield  {author} {\bibinfo {author} {\bibfnamefont {D.}~\bibnamefont
  {Miravet}}, \bibinfo {author} {\bibfnamefont {A.}~\bibnamefont
  {Alt{\i}nta{\c{s}}}}, \bibinfo {author} {\bibfnamefont {A.~W.}\ \bibnamefont
  {Rodrigues}}, \bibinfo {author} {\bibfnamefont {M.}~\bibnamefont {Bieniek}},
  \bibinfo {author} {\bibfnamefont {M.}~\bibnamefont {Korkusinski}},\ and\
  \bibinfo {author} {\bibfnamefont {P.}~\bibnamefont {Hawrylak}},\ }\href
  {https://doi.org/https://doi.org/10.1103/PhysRevB.108.195407} {\bibfield
  {journal} {\bibinfo  {journal} {Phys. Rev. B}\ }\textbf {\bibinfo {volume}
  {108}},\ \bibinfo {pages} {195407} (\bibinfo {year} {2023})}\BibitemShut
  {NoStop}%
\bibitem [{\citenamefont {Mitra}\ \emph {et~al.}(2023)\citenamefont {Mitra},
  \citenamefont {Zafar},\ and\ \citenamefont {Apalkov}}]{vadymqd}%
  \BibitemOpen
  \bibfield  {author} {\bibinfo {author} {\bibfnamefont {A.}~\bibnamefont
  {Mitra}}, \bibinfo {author} {\bibfnamefont {A.~J.}\ \bibnamefont {Zafar}},\
  and\ \bibinfo {author} {\bibfnamefont {V.}~\bibnamefont {Apalkov}},\
  }\bibfield  {journal} {\bibinfo  {journal} {J. Condens. Matter Phys.}\ }\href
  {https://doi.org/10.1088/1361-648x/ad271a} {10.1088/1361-648x/ad271a}
  (\bibinfo {year} {2023})\BibitemShut {NoStop}%
\bibitem [{\citenamefont {Mitra}\ \emph {et~al.}(2024)\citenamefont {Mitra},
  \citenamefont {Zafar}, \citenamefont {Hosseini},\ and\ \citenamefont
  {Apalkov}}]{vadym2024ultrafast}%
  \BibitemOpen
  \bibfield  {author} {\bibinfo {author} {\bibfnamefont {A.}~\bibnamefont
  {Mitra}}, \bibinfo {author} {\bibfnamefont {A.~J.}\ \bibnamefont {Zafar}},
  \bibinfo {author} {\bibfnamefont {S.~J.}\ \bibnamefont {Hosseini}},\ and\
  \bibinfo {author} {\bibfnamefont {V.}~\bibnamefont {Apalkov}},\ }\href
  {https://doi.org/https://doi.org/10.1103/PhysRevB.109.155425} {\bibfield
  {journal} {\bibinfo  {journal} {Phys. Rev. B}\ }\textbf {\bibinfo {volume}
  {109}},\ \bibinfo {pages} {155425} (\bibinfo {year} {2024})}\BibitemShut
  {NoStop}%
\bibitem [{\citenamefont {Yannouleas}\ and\ \citenamefont
  {Landman}(2023)}]{2023quantWigner}%
  \BibitemOpen
  \bibfield  {author} {\bibinfo {author} {\bibfnamefont {C.}~\bibnamefont
  {Yannouleas}}\ and\ \bibinfo {author} {\bibfnamefont {U.}~\bibnamefont
  {Landman}},\ }\href
  {https://doi.org/https://doi.org/10.1103/PhysRevB.108.L121411} {\bibfield
  {journal} {\bibinfo  {journal} {Phys. Rev. B}\ }\textbf {\bibinfo {volume}
  {108}},\ \bibinfo {pages} {L121411} (\bibinfo {year} {2023})}\BibitemShut
  {NoStop}%
\bibitem [{\citenamefont {Zerba}\ \emph {et~al.}(2024)\citenamefont {Zerba},
  \citenamefont {Kuhlenkamp}, \citenamefont {Imamo\u{g}lu},\ and\ \citenamefont
  {Knap}}]{topsuper2024}%
  \BibitemOpen
  \bibfield  {author} {\bibinfo {author} {\bibfnamefont {C.}~\bibnamefont
  {Zerba}}, \bibinfo {author} {\bibfnamefont {C.}~\bibnamefont {Kuhlenkamp}},
  \bibinfo {author} {\bibfnamefont {A.}~\bibnamefont {Imamo\u{g}lu}},\ and\
  \bibinfo {author} {\bibfnamefont {M.}~\bibnamefont {Knap}},\ }\href
  {https://doi.org/10.1103/PhysRevLett.133.056902} {\bibfield  {journal}
  {\bibinfo  {journal} {Phys. Rev. Lett.}\ }\textbf {\bibinfo {volume} {133}},\
  \bibinfo {pages} {056902} (\bibinfo {year} {2024})}\BibitemShut {NoStop}%
\bibitem [{\citenamefont {Li}\ \emph {et~al.}(2015)\citenamefont {Li},
  \citenamefont {Zhong},\ and\ \citenamefont {Zhang}}]{hbn71}%
  \BibitemOpen
  \bibfield  {author} {\bibinfo {author} {\bibfnamefont {J.}~\bibnamefont
  {Li}}, \bibinfo {author} {\bibfnamefont {Y.}~\bibnamefont {Zhong}},\ and\
  \bibinfo {author} {\bibfnamefont {D.}~\bibnamefont {Zhang}},\ }\href
  {https://doi.org/10.1088/0953-8984/27/31/315301} {\bibfield  {journal}
  {\bibinfo  {journal} {J. Condens. Matter Phys.}\ }\textbf {\bibinfo {volume}
  {27}},\ \bibinfo {pages} {315301} (\bibinfo {year} {2015})}\BibitemShut
  {NoStop}%
\bibitem [{\citenamefont {Luo}\ \emph {et~al.}(2023)\citenamefont {Luo},
  \citenamefont {Reddy}, \citenamefont {Devakul},\ and\ \citenamefont
  {Fu}}]{luo2023artificial}%
  \BibitemOpen
  \bibfield  {author} {\bibinfo {author} {\bibfnamefont {D.}~\bibnamefont
  {Luo}}, \bibinfo {author} {\bibfnamefont {A.~P.}\ \bibnamefont {Reddy}},
  \bibinfo {author} {\bibfnamefont {T.}~\bibnamefont {Devakul}},\ and\ \bibinfo
  {author} {\bibfnamefont {L.}~\bibnamefont {Fu}},\ }\bibfield  {journal}
  {\bibinfo  {journal} {arXiv:2303.08162}\ }\href
  {https://doi.org/https://doi.org/10.48550/arXiv.2303.08162}
  {https://doi.org/10.48550/arXiv.2303.08162} (\bibinfo {year}
  {2023})\BibitemShut {NoStop}%
\bibitem [{\citenamefont {Berry}\ and\ \citenamefont
  {Mondragon}(1987)}]{berry1987neutrino}%
  \BibitemOpen
  \bibfield  {author} {\bibinfo {author} {\bibfnamefont {M.~V.}\ \bibnamefont
  {Berry}}\ and\ \bibinfo {author} {\bibfnamefont {R.}~\bibnamefont
  {Mondragon}},\ }\href
  {https://doi.org/https://doi.org/10.1098/rspa.1987.0080} {\bibfield
  {journal} {\bibinfo  {journal} {Proc. Roy. Soc. London A}\ }\textbf {\bibinfo
  {volume} {412}},\ \bibinfo {pages} {53} (\bibinfo {year} {1987})}\BibitemShut
  {NoStop}%
\bibitem [{\citenamefont {Gruji{\'c}}\ \emph {et~al.}(2011)\citenamefont
  {Gruji{\'c}}, \citenamefont {Zarenia}, \citenamefont {Chaves}, \citenamefont
  {Tadi{\'c}}, \citenamefont {Farias},\ and\ \citenamefont
  {Peeters}}]{pe2011electronic}%
  \BibitemOpen
  \bibfield  {author} {\bibinfo {author} {\bibfnamefont {M.}~\bibnamefont
  {Gruji{\'c}}}, \bibinfo {author} {\bibfnamefont {M.}~\bibnamefont {Zarenia}},
  \bibinfo {author} {\bibfnamefont {A.}~\bibnamefont {Chaves}}, \bibinfo
  {author} {\bibfnamefont {M.}~\bibnamefont {Tadi{\'c}}}, \bibinfo {author}
  {\bibfnamefont {G.}~\bibnamefont {Farias}},\ and\ \bibinfo {author}
  {\bibfnamefont {F.}~\bibnamefont {Peeters}},\ }\href
  {https://doi.org/https://doi.org/10.1103/PhysRevB.84.205441} {\bibfield
  {journal} {\bibinfo  {journal} {Phys. Rev. B}\ }\textbf {\bibinfo {volume}
  {84}},\ \bibinfo {pages} {205441} (\bibinfo {year} {2011})}\BibitemShut
  {NoStop}%
\bibitem [{\citenamefont {Paananen}\ and\ \citenamefont
  {Egger}(2011)}]{egg2011finite}%
  \BibitemOpen
  \bibfield  {author} {\bibinfo {author} {\bibfnamefont {T.}~\bibnamefont
  {Paananen}}\ and\ \bibinfo {author} {\bibfnamefont {R.}~\bibnamefont
  {Egger}},\ }\href
  {https://doi.org/https://doi.org/10.1103/PhysRevB.84.155456} {\bibfield
  {journal} {\bibinfo  {journal} {Phys. Rev. B}\ }\textbf {\bibinfo {volume}
  {84}},\ \bibinfo {pages} {155456} (\bibinfo {year} {2011})}\BibitemShut
  {NoStop}%
\bibitem [{\citenamefont {Paananen}\ \emph {et~al.}(2011)\citenamefont
  {Paananen}, \citenamefont {Egger},\ and\ \citenamefont
  {Siedentop}}]{egger2011signatures}%
  \BibitemOpen
  \bibfield  {author} {\bibinfo {author} {\bibfnamefont {T.}~\bibnamefont
  {Paananen}}, \bibinfo {author} {\bibfnamefont {R.}~\bibnamefont {Egger}},\
  and\ \bibinfo {author} {\bibfnamefont {H.}~\bibnamefont {Siedentop}},\ }\href
  {https://doi.org/https://doi.org/10.1103/PhysRevB.83.085409} {\bibfield
  {journal} {\bibinfo  {journal} {Phys. Rev. B}\ }\textbf {\bibinfo {volume}
  {83}},\ \bibinfo {pages} {085409} (\bibinfo {year} {2011})}\BibitemShut
  {NoStop}%
\bibitem [{\citenamefont {Raca}\ and\ \citenamefont
  {Milovanovi{\'c}}(2017)}]{dirac2017excitonic}%
  \BibitemOpen
  \bibfield  {author} {\bibinfo {author} {\bibfnamefont {V.}~\bibnamefont
  {Raca}}\ and\ \bibinfo {author} {\bibfnamefont {M.}~\bibnamefont
  {Milovanovi{\'c}}},\ }\href
  {https://doi.org/https://doi.org/10.1103/PhysRevB.96.195434} {\bibfield
  {journal} {\bibinfo  {journal} {Phys. Rev. B}\ }\textbf {\bibinfo {volume}
  {96}},\ \bibinfo {pages} {195434} (\bibinfo {year} {2017})}\BibitemShut
  {NoStop}%
\bibitem [{\citenamefont {Zhou}\ \emph {et~al.}(2023)\citenamefont {Zhou},
  \citenamefont {Li}, \citenamefont {Zhang}, \citenamefont {Wang},
  \citenamefont {Fan}, \citenamefont {Zou}, \citenamefont {Cai}, \citenamefont
  {Jiang}, \citenamefont {Zhou}, \citenamefont {Zhang} \emph
  {et~al.}}]{zhou2023controllable}%
  \BibitemOpen
  \bibfield  {author} {\bibinfo {author} {\bibfnamefont {Y.}~\bibnamefont
  {Zhou}}, \bibinfo {author} {\bibfnamefont {C.}~\bibnamefont {Li}}, \bibinfo
  {author} {\bibfnamefont {Y.}~\bibnamefont {Zhang}}, \bibinfo {author}
  {\bibfnamefont {L.}~\bibnamefont {Wang}}, \bibinfo {author} {\bibfnamefont
  {X.}~\bibnamefont {Fan}}, \bibinfo {author} {\bibfnamefont {L.}~\bibnamefont
  {Zou}}, \bibinfo {author} {\bibfnamefont {Z.}~\bibnamefont {Cai}}, \bibinfo
  {author} {\bibfnamefont {J.}~\bibnamefont {Jiang}}, \bibinfo {author}
  {\bibfnamefont {S.}~\bibnamefont {Zhou}}, \bibinfo {author} {\bibfnamefont
  {B.}~\bibnamefont {Zhang}}, \emph {et~al.},\ }\href
  {https://doi.org/https://doi.org/10.1002/adfm.202304302} {\bibfield
  {journal} {\bibinfo  {journal} {Adv. Funct. Mater.}\ }\textbf {\bibinfo
  {volume} {33}},\ \bibinfo {pages} {2304302} (\bibinfo {year}
  {2023})}\BibitemShut {NoStop}%
\bibitem [{\citenamefont {Chu}\ \emph {et~al.}(2024)\citenamefont {Chu},
  \citenamefont {Zhou}, \citenamefont {Wang}, \citenamefont {Fan},
  \citenamefont {Guo}, \citenamefont {Li}, \citenamefont {Yue}, \citenamefont
  {Ouyang}, \citenamefont {Zhao},\ and\ \citenamefont {Zhou}}]{zhou2024stable}%
  \BibitemOpen
  \bibfield  {author} {\bibinfo {author} {\bibfnamefont {W.}~\bibnamefont
  {Chu}}, \bibinfo {author} {\bibfnamefont {X.}~\bibnamefont {Zhou}}, \bibinfo
  {author} {\bibfnamefont {Z.}~\bibnamefont {Wang}}, \bibinfo {author}
  {\bibfnamefont {X.}~\bibnamefont {Fan}}, \bibinfo {author} {\bibfnamefont
  {X.}~\bibnamefont {Guo}}, \bibinfo {author} {\bibfnamefont {C.}~\bibnamefont
  {Li}}, \bibinfo {author} {\bibfnamefont {J.}~\bibnamefont {Yue}}, \bibinfo
  {author} {\bibfnamefont {F.}~\bibnamefont {Ouyang}}, \bibinfo {author}
  {\bibfnamefont {J.}~\bibnamefont {Zhao}},\ and\ \bibinfo {author}
  {\bibfnamefont {Y.}~\bibnamefont {Zhou}},\ }\href
  {https://doi.org/10.1007/s11467-024-1414-7} {\bibfield  {journal} {\bibinfo
  {journal} {Front. Phys.}\ }\textbf {\bibinfo {volume} {19}},\ \bibinfo
  {pages} {1} (\bibinfo {year} {2024})}\BibitemShut {NoStop}%
\bibitem [{\citenamefont {Keum}\ \emph {et~al.}(2015)\citenamefont {Keum},
  \citenamefont {Cho}, \citenamefont {Kim}, \citenamefont {Choe}, \citenamefont
  {Sung}, \citenamefont {Kan}, \citenamefont {Kang}, \citenamefont {Hwang},
  \citenamefont {Kim}, \citenamefont {Yang} \emph {et~al.}}]{MoTe2phase}%
  \BibitemOpen
  \bibfield  {author} {\bibinfo {author} {\bibfnamefont {D.~H.}\ \bibnamefont
  {Keum}}, \bibinfo {author} {\bibfnamefont {S.}~\bibnamefont {Cho}}, \bibinfo
  {author} {\bibfnamefont {J.~H.}\ \bibnamefont {Kim}}, \bibinfo {author}
  {\bibfnamefont {D.-H.}\ \bibnamefont {Choe}}, \bibinfo {author}
  {\bibfnamefont {H.-J.}\ \bibnamefont {Sung}}, \bibinfo {author}
  {\bibfnamefont {M.}~\bibnamefont {Kan}}, \bibinfo {author} {\bibfnamefont
  {H.}~\bibnamefont {Kang}}, \bibinfo {author} {\bibfnamefont {J.-Y.}\
  \bibnamefont {Hwang}}, \bibinfo {author} {\bibfnamefont {S.~W.}\ \bibnamefont
  {Kim}}, \bibinfo {author} {\bibfnamefont {H.}~\bibnamefont {Yang}}, \emph
  {et~al.},\ }\href {https://doi.org/https://doi.org/10.1038/nphys3314}
  {\bibfield  {journal} {\bibinfo  {journal} {Nat. Phys.}\ }\textbf {\bibinfo
  {volume} {11}},\ \bibinfo {pages} {482} (\bibinfo {year} {2015})}\BibitemShut
  {NoStop}%
\bibitem [{\citenamefont {Tongay}\ \emph {et~al.}(2014)\citenamefont {Tongay},
  \citenamefont {Sahin}, \citenamefont {Ko}, \citenamefont {Luce},
  \citenamefont {Fan}, \citenamefont {Liu}, \citenamefont {Zhou}, \citenamefont
  {Huang}, \citenamefont {Ho}, \citenamefont {Yan} \emph {et~al.}}]{bulkReS2}%
  \BibitemOpen
  \bibfield  {author} {\bibinfo {author} {\bibfnamefont {S.}~\bibnamefont
  {Tongay}}, \bibinfo {author} {\bibfnamefont {H.}~\bibnamefont {Sahin}},
  \bibinfo {author} {\bibfnamefont {C.}~\bibnamefont {Ko}}, \bibinfo {author}
  {\bibfnamefont {A.}~\bibnamefont {Luce}}, \bibinfo {author} {\bibfnamefont
  {W.}~\bibnamefont {Fan}}, \bibinfo {author} {\bibfnamefont {K.}~\bibnamefont
  {Liu}}, \bibinfo {author} {\bibfnamefont {J.}~\bibnamefont {Zhou}}, \bibinfo
  {author} {\bibfnamefont {Y.-S.}\ \bibnamefont {Huang}}, \bibinfo {author}
  {\bibfnamefont {C.-H.}\ \bibnamefont {Ho}}, \bibinfo {author} {\bibfnamefont
  {J.}~\bibnamefont {Yan}}, \emph {et~al.},\ }\href
  {https://doi.org/https://doi.org/10.1038/ncomms4252} {\bibfield  {journal}
  {\bibinfo  {journal} {Nat. Commun.}\ }\textbf {\bibinfo {volume} {5}},\
  \bibinfo {pages} {3252} (\bibinfo {year} {2014})}\BibitemShut {NoStop}%
\bibitem [{\citenamefont {Pietil{\"a}inen}\ and\ \citenamefont
  {Chakraborty}(2006)}]{tapsh2006energy}%
  \BibitemOpen
  \bibfield  {author} {\bibinfo {author} {\bibfnamefont {P.}~\bibnamefont
  {Pietil{\"a}inen}}\ and\ \bibinfo {author} {\bibfnamefont {T.}~\bibnamefont
  {Chakraborty}},\ }\href
  {https://doi.org/https://doi.org/10.1103/PhysRevB.73.155315} {\bibfield
  {journal} {\bibinfo  {journal} {Phys. Rev. B}\ }\textbf {\bibinfo {volume}
  {73}},\ \bibinfo {pages} {155315} (\bibinfo {year} {2006})}\BibitemShut
  {NoStop}%
\bibitem [{\citenamefont {Wang}\ \emph {et~al.}(2023)\citenamefont {Wang},
  \citenamefont {Sedrakyan}, \citenamefont {Wang}, \citenamefont {Du},\ and\
  \citenamefont {Du}}]{du2023fqhe}%
  \BibitemOpen
  \bibfield  {author} {\bibinfo {author} {\bibfnamefont {R.}~\bibnamefont
  {Wang}}, \bibinfo {author} {\bibfnamefont {T.~A.}\ \bibnamefont {Sedrakyan}},
  \bibinfo {author} {\bibfnamefont {B.}~\bibnamefont {Wang}}, \bibinfo {author}
  {\bibfnamefont {L.}~\bibnamefont {Du}},\ and\ \bibinfo {author}
  {\bibfnamefont {R.-R.}\ \bibnamefont {Du}},\ }\href
  {https://doi.org/https://doi.org/10.1038/s41586-023-06065-w} {\bibfield
  {journal} {\bibinfo  {journal} {Nature}\ }\textbf {\bibinfo {volume} {619}},\
  \bibinfo {pages} {57} (\bibinfo {year} {2023})}\BibitemShut {NoStop}%
\bibitem [{\citenamefont {Camenzind}\ \emph {et~al.}(2019)\citenamefont
  {Camenzind}, \citenamefont {Yu}, \citenamefont {Stano}, \citenamefont
  {Zimmerman}, \citenamefont {Gossard}, \citenamefont {Loss},\ and\
  \citenamefont {Zumb{\"u}hl}}]{camenzind2019spectroscopy}%
  \BibitemOpen
  \bibfield  {author} {\bibinfo {author} {\bibfnamefont {L.~C.}\ \bibnamefont
  {Camenzind}}, \bibinfo {author} {\bibfnamefont {L.}~\bibnamefont {Yu}},
  \bibinfo {author} {\bibfnamefont {P.}~\bibnamefont {Stano}}, \bibinfo
  {author} {\bibfnamefont {J.~D.}\ \bibnamefont {Zimmerman}}, \bibinfo {author}
  {\bibfnamefont {A.~C.}\ \bibnamefont {Gossard}}, \bibinfo {author}
  {\bibfnamefont {D.}~\bibnamefont {Loss}},\ and\ \bibinfo {author}
  {\bibfnamefont {D.~M.}\ \bibnamefont {Zumb{\"u}hl}},\ }\href
  {https://doi.org/https://doi.org/10.1103/PhysRevLett.122.207701} {\bibfield
  {journal} {\bibinfo  {journal} {Phys. Rev. Lett.}\ }\textbf {\bibinfo
  {volume} {122}},\ \bibinfo {pages} {207701} (\bibinfo {year}
  {2019})}\BibitemShut {NoStop}%
\bibitem [{\citenamefont {Luo}\ \emph {et~al.}(2024)\citenamefont {Luo},
  \citenamefont {Peng},\ and\ \citenamefont
  {Chakraborty}}]{luo2024encyclopedia}%
  \BibitemOpen
  \bibfield  {author} {\bibinfo {author} {\bibfnamefont {W.}~\bibnamefont
  {Luo}}, \bibinfo {author} {\bibfnamefont {S.}~\bibnamefont {Peng}},\ and\
  \bibinfo {author} {\bibfnamefont {T.}~\bibnamefont {Chakraborty}},\ }in\
  \href {https://doi.org/https://doi.org/10.1016/B978-0-323-90800-9.00046-9}
  {\emph {\bibinfo {booktitle} {Encyclopedia of Condensed Matter Physics}}},\
  \bibinfo {editor} {edited by\ \bibinfo {editor} {\bibfnamefont
  {T.}~\bibnamefont {Chakraborty}}}\ (\bibinfo  {publisher} {Academic Press},\
  \bibinfo {year} {2024})\ \bibinfo {edition} {2nd}\ ed.,\ pp.\ \bibinfo
  {pages} {400--414}\BibitemShut {NoStop}%
\bibitem [{\citenamefont {Luo}\ and\ \citenamefont
  {Chakraborty}(2019)}]{luo2019tuning}%
  \BibitemOpen
  \bibfield  {author} {\bibinfo {author} {\bibfnamefont {W.}~\bibnamefont
  {Luo}}\ and\ \bibinfo {author} {\bibfnamefont {T.}~\bibnamefont
  {Chakraborty}},\ }\href
  {https://doi.org/https://doi.org/10.1103/PhysRevB.100.085309} {\bibfield
  {journal} {\bibinfo  {journal} {Phys. Rev. B}\ }\textbf {\bibinfo {volume}
  {100}},\ \bibinfo {pages} {085309} (\bibinfo {year} {2019})}\BibitemShut
  {NoStop}%
\bibitem [{\citenamefont {Avetisyan}\ \emph {et~al.}(2012)\citenamefont
  {Avetisyan}, \citenamefont {Pietil{\"a}inen},\ and\ \citenamefont
  {Chakraborty}}]{avetisyan2012strong}%
  \BibitemOpen
  \bibfield  {author} {\bibinfo {author} {\bibfnamefont {S.}~\bibnamefont
  {Avetisyan}}, \bibinfo {author} {\bibfnamefont {P.}~\bibnamefont
  {Pietil{\"a}inen}},\ and\ \bibinfo {author} {\bibfnamefont {T.}~\bibnamefont
  {Chakraborty}},\ }\href
  {https://doi.org/https://doi.org/10.1103/PhysRevB.85.153301} {\bibfield
  {journal} {\bibinfo  {journal} {Phys. Rev. B}\ }\textbf {\bibinfo {volume}
  {85}},\ \bibinfo {pages} {153301} (\bibinfo {year} {2012})}\BibitemShut
  {NoStop}%
\bibitem [{\citenamefont {Avetisyan}\ \emph {et~al.}(2013)\citenamefont
  {Avetisyan}, \citenamefont {Pietil{\"a}inen},\ and\ \citenamefont
  {Chakraborty}}]{avetisyan2013superintense}%
  \BibitemOpen
  \bibfield  {author} {\bibinfo {author} {\bibfnamefont {S.}~\bibnamefont
  {Avetisyan}}, \bibinfo {author} {\bibfnamefont {P.}~\bibnamefont
  {Pietil{\"a}inen}},\ and\ \bibinfo {author} {\bibfnamefont {T.}~\bibnamefont
  {Chakraborty}},\ }\href
  {https://doi.org/https://doi.org/10.1103/PhysRevB.88.205310} {\bibfield
  {journal} {\bibinfo  {journal} {Phys. Rev. B}\ }\textbf {\bibinfo {volume}
  {88}},\ \bibinfo {pages} {205310} (\bibinfo {year} {2013})}\BibitemShut
  {NoStop}%
\bibitem [{\citenamefont {Chakraborty}\ \emph
  {et~al.}(2018{\natexlab{b}})\citenamefont {Chakraborty}, \citenamefont
  {Manaselyan}, \citenamefont {Barseghyan},\ and\ \citenamefont
  {Laroze}}]{chakraborty2018controllable}%
  \BibitemOpen
  \bibfield  {author} {\bibinfo {author} {\bibfnamefont {T.}~\bibnamefont
  {Chakraborty}}, \bibinfo {author} {\bibfnamefont {A.}~\bibnamefont
  {Manaselyan}}, \bibinfo {author} {\bibfnamefont {M.}~\bibnamefont
  {Barseghyan}},\ and\ \bibinfo {author} {\bibfnamefont {D.}~\bibnamefont
  {Laroze}},\ }\href
  {https://doi.org/https://doi.org/10.1103/PhysRevB.97.041304} {\bibfield
  {journal} {\bibinfo  {journal} {Phys. Rev. B}\ }\textbf {\bibinfo {volume}
  {97}},\ \bibinfo {pages} {041304} (\bibinfo {year}
  {2018}{\natexlab{b}})}\BibitemShut {NoStop}%
\end{thebibliography}%
\end{document}